# Optically pumped spin polarization as a probe of many-body thermalization

Daniela Pagliero[1,*], Pablo R. Zangara[6,7,*], Jacob Henshaw[1], Ashok Ajoy[3], Rodolfo H. Acosta[6,7], Jeffrey A. Reimer[4,5], Alexander Pines[3,5], Carlos A. Meriles[1,2,†]

[1]Department. of Physics, CUNY-City College of New York, New York, NY 10031, USA. [2]CUNY-Graduate Center, New York, NY 10016, USA. [3]Department of Chemistry, University of California at Berkeley, Berkeley, California 94720, USA. [4]Department of Chemical and Biomolecular Engineering, University of California at Berkeley, Berkeley, California 94720, USA. [5]Materials Science Division Lawrence Berkeley National Laboratory, Berkeley, California 94720, USA. [6]Facultad de Matemática, Astronomía, Física y Computación, Universidad Nacional de Córdoba, Ciudad Universitaria, CP:X5000HUA Córdoba, Argentina. [7]Instituto de Física Enrique Gaviola (IFEG), CONICET, Medina Allende s/n, X5000HUA, Córdoba, Argentina. [*]Equally contributing authors. [†]Corresponding author. E-mail: cmeriles@ccny.cuny.edu

The interplay between disorder and transport is a problem central to the understanding of a broad range of physical processes, most notably the ability of a system to reach thermal equilibrium. Disorder and many body interactions are known to compete, with the dominance of one or the other giving rise to fundamentally different dynamical phases. Here we investigate the spin diffusion dynamics of $^{13}$C in diamond, which we dynamically polarize at room temperature via optical spin pumping of engineered color centers. We focus on low-abundance, strongly hyperfine-coupled nuclei, whose role in the polarization transport we expose through the integrated impact of variable radio-frequency excitation on the observable bulk $^{13}$C magnetic resonance signal. Unexpectedly, we find good thermal contact throughout the nuclear spin bath, virtually independent of the hyperfine coupling strength, which we attribute to effective carbon-carbon interactions mediated by the electronic spin ensemble. In particular, observations across the full range of hyperfine couplings indicate the nuclear spin diffusion constant takes values up to two orders of magnitude greater than that expected from homo-nuclear spin couplings. Our results open intriguing opportunities to study the onset of thermalization in a system by controlling the internal interactions within the bath.

## INTRODUCTION

Although the quest to understand the roles of disorder and couplings in the out-of-equilibrium dynamics of many body systems goes back several decades[1], the field is presently witnessing a resurgence, in part due to its intrinsic connection to the development of novel quantum technologies. Progress has been made largely possible by captivating experiments in cold gases where the coupling to outer reservoirs can be virtually suppressed and the evolution of each of the atoms in the interacting ensemble is probed individually[2-4]. An example of recently observed phenomena is many-body localization[5,6] (MBL), a process where, despite the interactions between its inner units, the system fails to thermalize, i.e., its long-term properties cannot be captured by conventional equilibrium statistical mechanics[7,8]. Unlike Anderson-localization[9], inter-particle couplings lead to dephasing of individual, initially-localized states[7]. Interestingly, however, the absence of exchange between different MBL modes endows these systems with a long-term memory, which makes them potentially useful platforms to store and retrieve quantum information.

Interacting spins in diamond provide an intriguing platform to investigate the interplay between localization and thermalization because electrons and nuclei feature species-specific interactions and concentrations that can be tuned and dynamically controlled. Hyperfine couplings with paramagnetic centers can take extreme values (exceeding hundreds of MHz for first shell carbons), while the low gyromagnetic ratio and natural abundance of $^{13}$C spins make homonuclear couplings orders of magnitude weaker (~100 Hz). Given our understanding of thermalization as a spin diffusion process, the large frequency mismatch between hyperfine-coupled and bulk nuclei immediately raises questions on the system's ability to reach equilibrium. This problem — paramount to interpreting nuclear spin-lattice relaxation[10] but equally relevant to carrier transport[9,11] — has been traditionally explained through the notion of a 'spin diffusion barrier', i.e., a virtual line in the space around a paramagnetic center separating 'a frozen core' of nuclei unable to communicate (i.e., 'flip-flop') with bulk spins[12-14].

Here, we combine optical excitation and nuclear magnetic resonance (NMR) at low magnetic fields to investigate the generation and transport of nuclear magnetization in a diamond crystal hosting nitrogen-vacancy (NV) centers. Formed by a substitutional nitrogen immediately adjacent to a vacancy, these spin-1 point defects polarize efficiently under green illumination, which can be exploited to dynamically polarize the $^{13}$C nuclei in the crystal. Working under 'energy matching' conditions — where NVs cross-relax with surrounding spin-1/2 nitrogen impurities or 'P1 centers' — we find that strongly hyperfine-coupled carbons can efficiently exchange polarization with bulk nuclei; this process is made possible by many body interactions involving electron and nuclear spins through mechanisms we formally capture via a nuclear-spin-only effective Hamiltonian. Further, we measure nuclear spin diffusion constants across a range of hyperfine couplings orders of magnitude greater than the nuclear Larmor frequency, and find values ~100-fold bigger than those possible via homonuclear couplings, a phenomenon we interpret in terms of electron-mediated interactions between distant carbons.

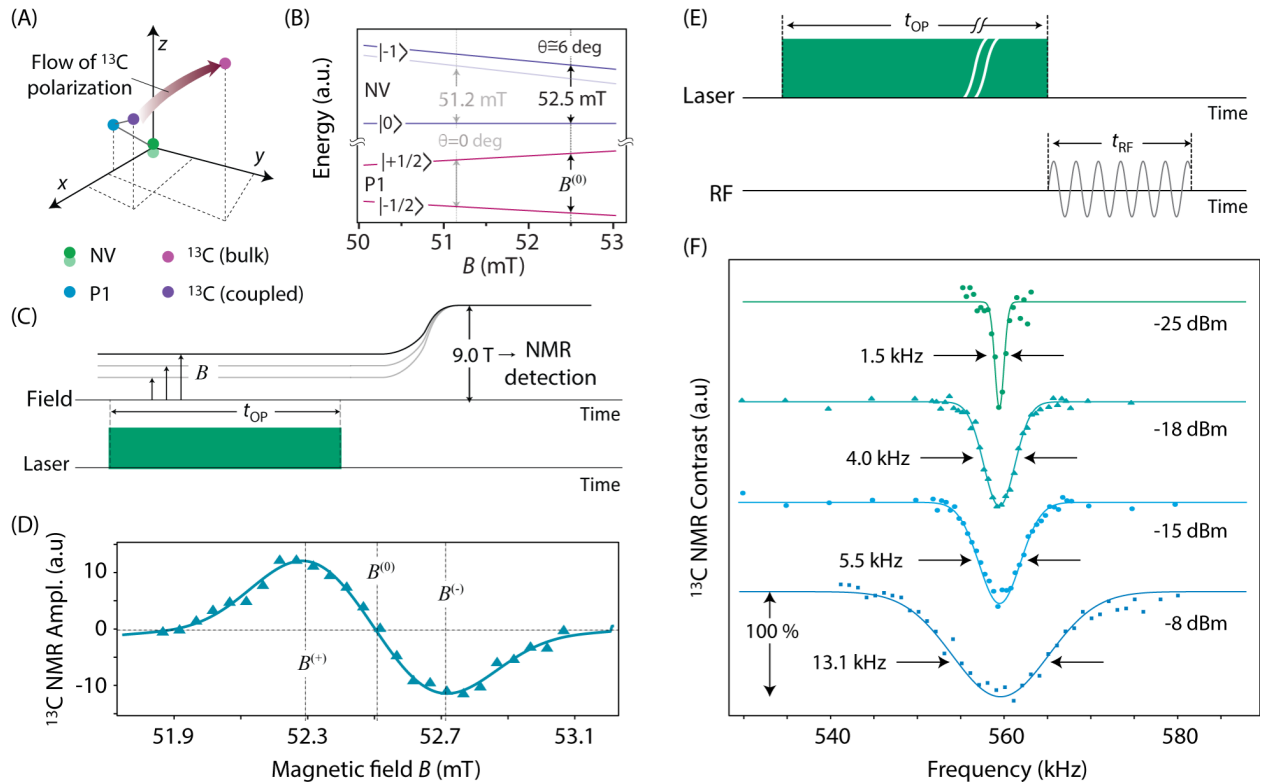

**Figure 1 | Low-field dynamic polarization and manipulation of $^{13}$C spins in diamond.** (A) Electron-nuclear spin set. Polarization flows from hyperfine-coupled carbons to bulk carbons. (B) Schematics of the NV/P1 energy diagrams as a function of the magnetic field. Cross-relaxation between the NV and P1 is most favorable when the energy differences are matched (vertical arrows); this condition depends on the angle θ between the magnetic field $B$ and the NV symmetry axis. (C) Dynamic nuclear polarization and detection protocol. We illuminate the sample with 532 nm laser light for a time $t_{OP}$ at a variable field $B$, followed by sample shuttling to the bore of a 9.0 T magnet for high-field $^{13}$C NMR detection. (D) NMR signal amplitude of hyperpolarized $^{13}$C as a function of $B$. In a typical experiment, the magnetic field during DNP is set at $B^{(+)}$ or at $B^{(-)}$, so as to produce the largest positive or negative $^{13}$C polarization, respectively. (E) Indirect observation of low-field $^{13}$C NMR through variable-frequency RF excitation; for simplicity, the drawing omits the sample shuttling step. (F) Experimental results from applying the protocol in (E) for different RF powers. In (D), (E) and (F), the optical pumping time is $t_{OP} = 10$ s and the laser power is 1 W focused to a ~200-μm-diameter focal spot; in (F) the RF-pulse duration is $t_{RF} = 250$ ms, the magnetic field is $B^{(+)} = 52.3$ mT, and its angle θ with the NV axis amounts to ~6 deg.

## RESULTS

### $^{13}$C hyperfine spectroscopy at low magnetic fields

Figs. 1A through 1C summarize the conditions in our experiments. We study a diamond sample with a large NV and P1 content (~10 and ~50 ppm, respectively) produced via high-energy electron irradiation and annealing. We operate in the regime of 'cross-relaxation' where the separation between the $|m_S = 0\rangle$ and $|m_S = -1\rangle$ energy levels of the NV approximately matches the P1 Zeeman splitting in an external magnetic field $B^{(0)}$, whose exact value depends on the angle θ with the NV axis[15,16]. Optical pumping of the NV induces dynamic nuclear polarization (DNP) of bulk $^{13}$C, which we subsequently detect using a field cycling protocol (Fig. 1C). Fig. 1D shows the amplitude of the observed $^{13}$C NMR signal as a function of the optical pumping field $B$: The DNP generation can be simplistically understood through an energy-conserving NV–P1–$^{13}$C process where nuclear spins polarize positively or negatively depending on the sign of the difference between the NV and P1 transitions above or below $B^{(0)}$. On the other hand, the fact that sizable DNP can be observed for a field mismatch as large as ~0.4 mT (corresponding to hyperfine couplings of order ~10 MHz) immediately points to non-trivial channels of polarization transfer from nuclear spins strongly coupled to defects.

To measure the $^{13}$C spectrum at a given optical pumping field, we apply a radio-frequency (RF) pulse immediately after laser illumination (prior to sample shuttling, Fig. 1E) within a range around the $^{13}$C Zeeman frequency. The pulse duration (1 s) is chosen so as to make the up/down $^{13}$C spin populations equal when on resonance, hence leading to a "dip" in the observed signal amplitudes plotted as a function of the RF frequency. Fig. 1F shows the results for variable RF power: In the limit of weak RF excitation (-25 dBm), the NMR linewidth amounts to ~1.5 kHz, coincident with that observed at high field (see Supplementary Material, Section I). Stronger RF power results in broader dips, a consequence of the greater excitation bandwidth; in the experiments below we use an RF power of -8 dBm, which confines the effect to a ~13 kHz band around

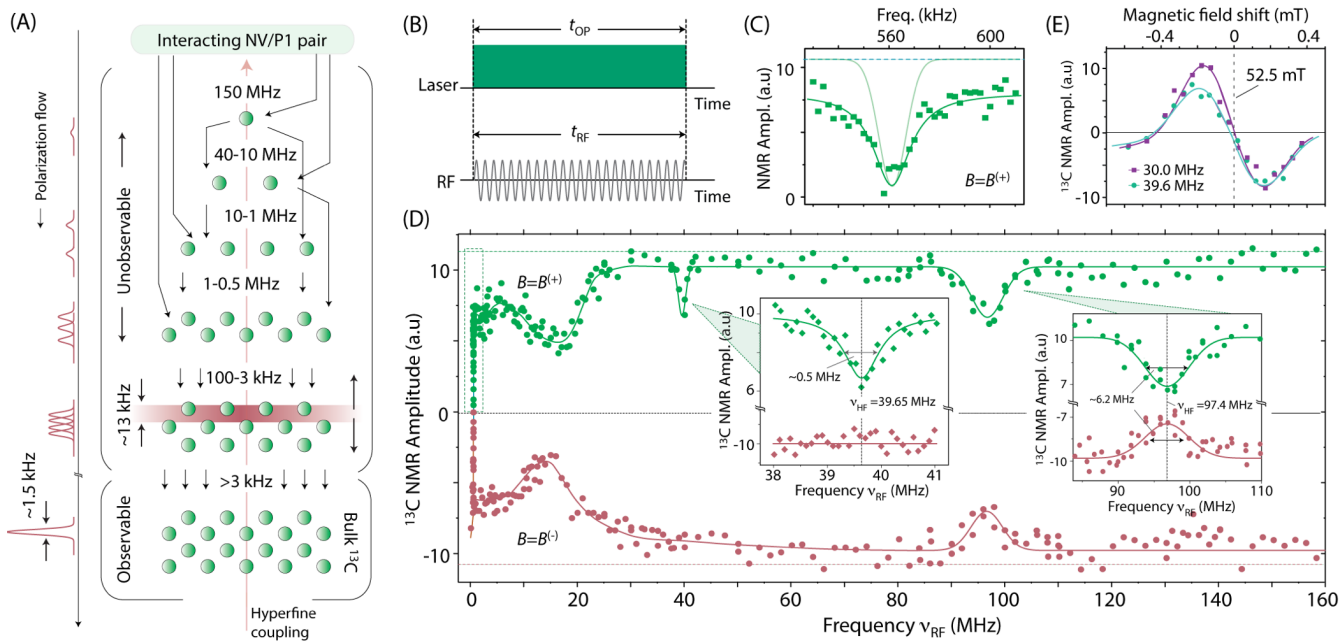

**Figure 2 | $^{13}$C spin diffusion spectroscopy via signal amplification of low-abundance nuclei.** (A) Schematics of the spin diffusion process. Starting with the cross-relaxation of an NV–P1 pair and a strongly-hyperfine-coupled $^{13}$C spin (green circles), polarization flows from less abundant, unobservable nuclei to more abundant, bulk carbons. RF excitation at a predefined (but variable) frequency equilibrates the populations of a select nuclear spin subset (horizontal red band), hence disrupting the polarization flow. (B) Experimental protocol. $^{13}$C NMR detection is carried out at 9.0 T, following sample shuttling (not shown). (C) $^{13}$C NMR signal amplitude as a function of the RF frequency upon application of the protocol in (B) in a vicinity of the $^{13}$C Larmor frequency at $B^{(+)} = 52.3$ mT. The faint solid trace reproduces the spectrum in Fig. 1F at 0 dBm. (D) Same as in (C) but for an extended RF range. Here, the magnetic field is $B^{(+)} = 52.3$ mT ($B^{(-)} = 52.7$ mT) in the upper (lower) half plot (green and red circles, respectively). The dashed green square on the left indicates the region of the spectrum presented in (C). Solid lines are guides to the eye; faint horizontal traces indicate signal levels in the absence of RF. (E) $^{13}$C NMR signal amplitude as a function of the applied magnetic field in the presence of RF excitation either resonant (39.6 MHz) or non-resonant (30.0 MHz) with the dip in (D). Solid lines are guides to eye. In (C), (D) and (E), the RF power is -8 dBm, and $t_{OP} = t_{RF} = 5$ s.

resonance.

The ability to manipulate $^{13}$C spins gives us the opportunity to probe the transport of spin magnetization from paramagnetic centers to 'bulk' (i.e., very weakly coupled) carbons as it cascades down across nuclear spins with different hyperfine couplings under NV–P1 cross relaxation. Given the multi-spin nature of the dynamics at work, this process is better visualized in frequency space as a sequence of jumps along a chain formed by groups of carbons with varying hyperfine coupling (and hence different resonance frequencies, Fig. 2A). Nuclear spins proximal to paramagnetic centers (NVs or P1s) are normally invisible in the standard NMR signal due to their comparatively low abundance and extreme hyperfine-induced gradients. Nevertheless, their ability to mediate the transfer of magnetization to bulk spins can be selectively exposed through the accumulated effect of RF excitation on the polarization buildup during optical spin pumping.

Initial evidence revealing the non-trivial role of hyperfine coupled carbons is shown in Fig. 2C where we compare the NMR signal amplitudes following simultaneous RF and laser excitation (Fig. 2B): Accompanying the expected dip near the $^{13}$C Zeeman transition ($\omega_I \sim 2\pi \times 560$ kHz), we observe (partial) NMR signal reduction over a wide frequency range (green squares), far exceeding the excitation bandwidth (faint, 13-kHz-broad Gaussian in the back here serving as a reference).

Intriguingly, we find this effect persists at even higher frequencies, where inter-carbon flip-flops should be strongly suppressed. This is further shown in Fig. 2D where we measure the equivalent of a hyperfine-resolved spectrum over a 160 MHz range, selectively sensitive to nuclear spins participating in the magnetization transport. We identify several high-frequency regions where RF excitation has a significant impact on the observed NMR signal, suggesting that localization — the regime naively anticipated for hyperfine-coupled carbons in a dilute nuclear spin system such as diamond — cannot capture the dynamics at play. Very much on the contrary, we show next that most nuclear spins communicate efficiently with each other despite their relatively large frequency mismatch.

To shed light on the underlying mechanisms, we start with a comparison between the RF absorption spectrum in Fig. 2D and the set of hyperfine couplings to NVs and P1s (respectively, colored bands in the background and vertical bars in Fig. 3A). We find a moderate correlation between the two: For example, the dip at ~40 MHz — associated to a second shell carbon around the P1 center[17] — suggests substitutional nitrogen plays an important role in enabling spin exchange between near-defect and bulk nuclei. Importantly, the dip disappears if one shifts the magnetic field from $B^{(+)}$ to $B^{(-)}$ — a change of only ~0.2 mT, see Fig. 2E — suggesting spin diffusion emerges from a multi-spin process requiring precise alignment between the NV, P1, and $^{13}$C energy levels. This notion is consistent with the very premise of DNP near ~51 mT, arising from nuclear-spin-assisted NV–P1 cross-relaxation at these fields[15,18]. On the other hand, one cannot rule out spin-lattice relaxation effects, as the bulk carbon $T_1$ time is also seen to moderately change, from ~5 s to ~7 s when transitioning from $B^{(+)}$ to $B^{(-)}$. Finally, the ~97.5 MHz resonance — which we could not match to any reported $^{13}$C site near the NV or P1 — may, instead, correspond to polarization pathways involving the nuclear spin of the $^{14}$N host at the P1 (known to participate in the polarization transfer[15,16,18]); additional work, however, will be needed to clarify its origin.

The absence of RF absorption is also an important indicator: For example, the flat response in Fig. 2D near $\nu_{RF}$~130 MHz — coincident with the hyperfine splitting of first shell carbons around the NV[19] — indicates these sites do not partake in the polarization transfer process, hence suggesting select nuclear spins — featuring exceedingly strong hyperfine interactions — fail to thermalize with the rest (see below). By the same token, no RF dips are observable between ~50 and ~90 MHz (omitted in Fig. 3A for simplicity), a range with no hyperfine coupled carbons[17,20,21].

More generally, the amplitude of the RF absorption dip reflects on the number of diffusion channels available to the system near a given excitation frequency $\nu_{RF}$ (Fig. 3B). A complete transport blockade — manifesting in the form of a full-contrast dip — is possible only when the nuclear spins resonant with the applied RF intervene in every polarization transfer event. As the number of alternative channels increases, the RF-induced contrast diminishes because most spin diffusion pathways do not involve resonant nuclei. The latter, of course, depends on the granularity of the frequency jumps $\delta\nu_d(\nu_{RF})$ characterizing the multi-spin configurational change during spin diffusion; greater RF absorption can be regained as $\delta\nu_d(\nu_{RF})$ becomes comparable to (or smaller than) the RF bandwidth $\delta\nu_b$ (~13 kHz in the present experiments). We believe this interplay is responsible for the DNP signal response below ~10 MHz, where the number of carbon sites with comparable hyperfine couplings (and thus the number of spin diffusion pathways) increases rapidly, while the nuclear spin energy difference $\delta\nu_d(\nu_{RF})$ in each jump gradually fades away. On a related note, a close inspection of Fig. 2D shows a slight offset relative to the signal amplitude observed in the absence of RF (faint horizontal lines). We presently ignore its origin but hypothesize it could stem from weak RF absorption between many body electron spin states (i.e., weakly allowed 'zero-quantum' transitions), which subsequently causes nuclear spin relaxation. Additional experiments, however, are mandatory to clarify this point.

**Effective Hamiltonian and spin diffusion dynamics**

Deriving from first principles a Hamiltonian that correctly reproduces the behavior of interacting electron and nuclear spin ensembles — a problem at the center of ongoing efforts[22] — remains a challenging task. Here, we capture the dynamics at play by considering a pair of carbons, each interacting with one of two P1 centers, which, in turn, couple dipolarly to each other (Fig. 3C). Focusing first on the 'hyperfine-dominated' regime (where $\|A_1\|\sim\|A_2\|>\mathcal{J}_d>\omega_I$), we find the polarization can flow from one carbon to the other with an effective rate $J_{\text{eff}}\sim\omega_I^2\mathcal{J}_d/(2\bar{A}^2)$, where $\mathcal{J}_d$ is the inter-electronic dipolar coupling constant, and $\bar{A}=(\|A_1\|+\|A_2\|)/2$ denotes the average hyperfine coupling. Though stemming from high-order virtual processes, $J_{\text{eff}}$ can reach sizable values when the electron spin concentration is sufficiently high. As an illustration, for an electron spin dipolar coupling $\langle\mathcal{J}_d/2\pi\rangle\sim1$ MHz (corresponding to a nitrogen concentration of ~10 ppm[23]), we obtain $J_{\text{eff}}/2\pi\sim1$ kHz for $\bar{A}/2\pi\sim10$ MHz.

While the above effective coupling allows most hyperfine-shifted nuclei to communicate, we also find that transport can be suppressed if the hyperfine shift difference $\delta A=\|A_1\|-\|A_2\|$ between the two carbons is large. More formally, we express the condition for delocalization as

$$\mathcal{J}_d \gtrsim \omega_I \bar{A}\delta A/(\bar{A}^2-\delta A^2), \qquad (1)$$

increasingly difficult to meet as $\bar{A}$ approaches $\delta A$ (i.e., when $\|A_1\|\gg\|A_2\|$, see Supplementary Material, Section III). This is likely the scenario for first shell carbons ($A$~130 MHz), separated from the rest by a large spectral gap (see Fig. 3A). For completeness, it is worth mentioning that in the 'dipolar dominated' regime (where $\mathcal{J}_d>A,\omega_I$), the effective nuclear spin coupling takes the form $J_{\text{eff}}\sim A^2/(4\mathcal{J}_d)$. This expression shows, as expected, vanishing interaction for nuclei decoupled from paramagnetic defects ($\mathcal{J}_d>\omega_I>A$), but it also suggests $J_{\text{eff}}$ can be quite strong, potentially exceeding 10 kHz in the narrow window where $\mathcal{J}_d>A>\omega_I$ (Supplementary Material, Sections II and III).

Importantly, $\mathcal{J}_d$-induced state mixing activates transitions at frequencies other than those expected for pure nuclear spin flips. This is shown in Figs. 3C and 3D, where we plot the calculated nuclear spin polarization in a $^{13}$C–P1–P1–$^{13}$C chain under continuous RF excitation assuming both carbons start from a polarized state (see also Supplementary Material, Section IV). When $\mathcal{J}_d\sim0$, the system absorbs selectively at the single nuclear spin hyperfine transitions. As $\mathcal{J}_d$ increases, however, new dips corresponding to simultaneous nuclear and electron spin flips emerge. Given the range of possible spatial configurations in disordered spin ensembles, RF excitation should therefore yield broad bands of less-than-optimal DNP crudely centered around the hyperfine transitions, in qualitative agreement with our observations.

From the above considerations, we surmise the ensemble of paramagnetic defects can be thought of as an underlying network providing the couplings required for nuclear spins to thermalize[24]; correspondingly, the spin Hamiltonian for a group of (otherwise non-interacting) $N_I$ carbon spins $\mathbf{I}_i$ takes the form (see Supplementary Material, Section III)

$$H_{\text{eff}} = \sum_i^{N_I} \omega_I^{(i)} I_i^z + \sum_{i>j}^{N_I} \left( J_{\text{eff},zz}^{(ij)} I_i^z I_j^z + J_{\text{eff},xy}^{(ij)} (I_i^+ I_j^- + I_i^- I_j^+) \right) \qquad (2)$$

where $\omega_I^{(i)}$ denotes the (electron-spin-dependent) local field at the $i$-th nuclear spin site, and $J_{\text{eff},zz}$, $J_{\text{eff},xy}$ represent effective electron-spin-mediated inter-nuclear couplings, which, in general, must be seen as functions of the applied magnetic field and electron spin concentration.

While chain-like systems are often integrable, added spatial dimensions break any underlying symmetry and typically render the dynamics chaotic. A realistic simulation of the system at hand requires, therefore, the use of multi-dimensional spin arrays, an increasingly challenging task as the number of nuclei grows. Here, we qualitatively test the dynamics of the Hamiltonian in Eq. (2) using a model nuclear spin set of 22 carbons in a Cayley tree geometry assuming only the central spin is initially polarized (Fig. 3E). The effective couplings between nuclear spins in different rings grow to the outside of the tree, so as to emulate the transition between the hyperfine-dominated (i.e., $A > \mathcal{J}_d > \omega_I$) and the dipolar-dominated (i.e., $\mathcal{J}_d > A > \omega_I$) regimes. It is known that the interplay between the terms linear and bilinear in $I^z$ (respectively corresponding to the local potential and interaction terms of the Hubbard Hamiltonian in a carrier transport picture[7]) may lead to MBL. To make the numerical problem tractable, we assume below that the flip-flop terms are dominant and thus the system is in an ergodic phase.

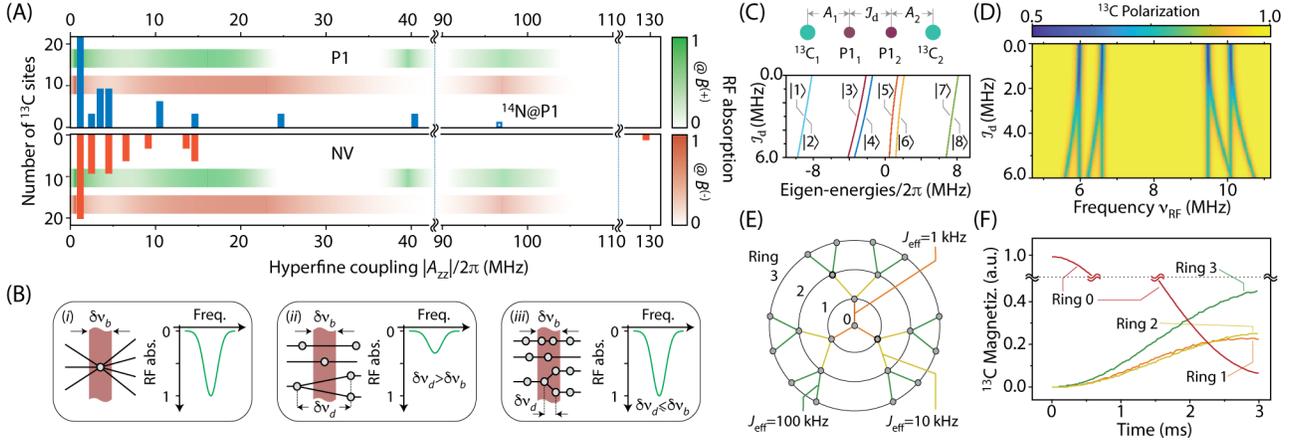

**Figure 3 | Electron-spin-mediated many-body nuclear spin diffusion under NV–P1 cross relaxation.** (A) Histograms of hyperfine resonance frequencies above 1 MHz for $^{13}$C nuclei near individual P1s and NVs (upper and lower plots, respectively). For reference, the faint green and red bands reproduce the level of RF absorption observed in Fig. 2D. (B) The impact of RF excitation on DNP efficiency can be cast in terms of a polarization sink of width $\delta\nu_b$ defined by the excited bandwidth. For a given RF power, the sink efficiency reflects on the spin network connectivity: (*i*) Full contrast arises when all polarization transfer pathways (solid lines) rely on a single nuclear spin site (grey circle) featuring a characteristic hyperfine shift. (*ii*) For a typical frequency change $\delta\nu_d$ between consecutive nuclear spin nodes and assuming $\delta\nu_d > \delta\nu_b$, the sink efficiency diminishes as the number of alternative pathways increases. (*iii*) Full contrast reappears when $\delta\nu_d \lesssim \delta\nu_b$. (C) (Top) Model spin chain comprising two carbons hyperfine-coupled to two P1s subject to a dipolar interaction $\mathcal{J}_d$. (Bottom) Calculated eigen-energies for eigen-states $|i\rangle$, $i = 1 \ldots 8$ within the subspace where the electron spins are anti-parallel; for these calculations, $\|A_1\| = 2\pi \times 6$ MHz, and $\|A_2\| = 2\pi \times 10$ MHz. (D) $^{13}$C polarization in the presence of RF for the spin system in (C) for different $\mathcal{J}_d$; both $^{13}$C spins are assumed initially polarized. (E) Network of 22 $^{13}$C spins in a Cayley tree configuration; green, yellow, and orange lines indicate $J_{\text{eff}}$ equal to 100 kHz, 10 kHz, and 1 kHz, respectively. (F) Computed $^{13}$C magnetization in each ring as a function of time starting from a configuration where only the central spin is polarized.

To compute the many body spin dynamics, we use a Trotter-Suzuki decomposition assisted by quantum parallelism[25]. Unlike other, more common approaches[26,27], this technique does not require truncation of the Hilbert space, and is thus applicable to long times (see Materials and Methods). As shown in Fig. 3F, we observe a diffusive (i.e., recurrence-free) evolution, pointing to the onset of quantum chaos[28-30]. Chaoticity arises in the Cayley geometry as a consequence of the system branching, effectively enlarging the size of the accessible Hilbert space as the polarization moves from inner to outer rings. Note that despite the growing inter-nuclear couplings, the characteristic time constant (of order ~2 ms) is uniform across the tree structure, dictated by the higher-order (and hence weaker) effective electronic couplings communicating the central spin with nuclei in the first ring.

Experimentally, we probe the time scale of spin diffusion in our sample via the protocol in Fig. 4A where we evenly distribute RF pulses of fixed duration throughout the illumination interval; the pulse length is chosen so as to ensure several $^{13}$C Rabi cycles (Supplementary Material, Section I). The upper half of Fig. 4B shows an example plot corresponding to radio-frequency at 10 MHz: For inter-pulse intervals $\tau \gtrsim 10$ ms, we find that the effect of RF pulses on the hyperpolarization amplitude is negligible, an early indication that spin diffusion takes place on a time scale faster than that deriving from direct inter-nuclear dipolar couplings (averaging ~100 Hz in non-enriched diamond). Overall, our data can be reasonably described via a stretched exponential dependence of the form $S = S_0 - S_1 \exp(-(\tau/\tau_d)^\varepsilon)$, where $\tau$ is the inter-pulse separation and all other variables are fitting parameters, with $\tau_d$ representing the characteristic nuclear spin diffusion time. Interestingly, we find $\varepsilon < 1$, typically indicative of heterogeneity in the underlying physical process[31,32]. This idea is consistent with the multi-channel nature of the transport dynamics at play, here expressed via the probability distribution $\mathcal{L}(\mu, \varepsilon)$ satisfying $\exp(-(\tau/\tau_d)^\varepsilon) = \int_0^\infty \mathcal{L}(\mu, \varepsilon) \exp(-\mu\tau) d\mu$. Using an inverse Laplace transform to explicitly compute $\mathcal{L}(\mu, \varepsilon)$, we find the distribution median satisfies $\bar{\mu} \sim 1/\tau_d$, i.e., diffusion rates are equally likely to lie above or below $1/\tau_d$ (lower half plot in Fig. 4B). In particular, we identify a broad set of fast transport processes whose rates extend beyond ~1 ms$^{-1}$ (shadowed tail in the plot).

To capture these observations into a functional microscopic model, we now return to the notion of magnetization transport along a one-dimensional (spectral) chain formed by $m$ spin sets $\{N_i\}$, $i: 1 \ldots m$ each featuring resonance frequencies within bands $\Delta\nu_i$ centered around effective hyperfine couplings $\|A_i\|$ (Fig. 4C). Aiming at a qualitative comparison with experiment, this time we model the transport problem classically using a set of coupled differential equations adapted to describe magnetization hops between boxes in the presence of optical and RF excitation, as well as nuclear spin-lattice relaxation (Supplementary Material, Section V). A train of RF pulses (resonant with nuclear spins within a band $\delta\nu_b$ along the chain) partly disrupts the transport of polarization and leads to a change $\delta M_m$ in the magnetization stored in box $m$ (here serving as the observable). This effect saturates in the limit where the inter-pulse separation $\tau$ is equal to (or shorter than) the interval required to replenish the magnetization in the depleted cell (roughly, the inflexion point in the sigmoid), hence allowing us to extract the characteristic spin diffusion time $\tau_d$ at the corresponding excitation frequency $\nu_{RF}$ (Fig. 4D).

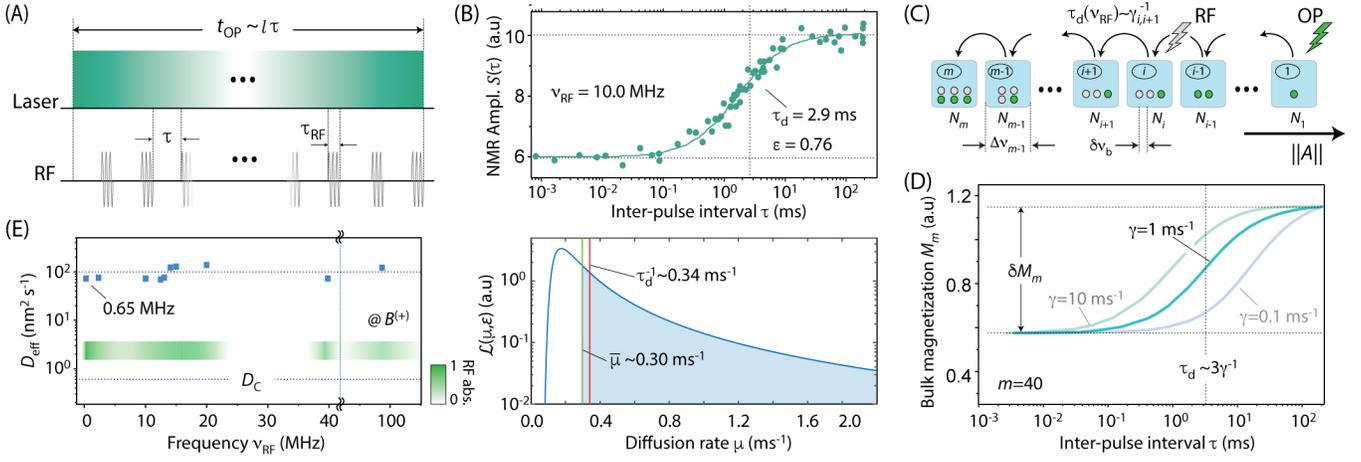

**Figure 4 | Probing paramagnetic-center-assisted nuclear spin diffusion.** (A) Experimental protocol. We apply a train of short, equidistant RF pulses during the fixed illumination time $t_{OP} = 5$ s and monitor the $^{13}$C DNP signal as we increase the number of pulses $l$. (B) (Top) $^{13}$C NMR signal amplitude $S(\tau)$ as a function of the inter-pulse time $\tau \approx t_{OP}/l$ at a representative radio-frequency. The RF pulse duration is $\tau_{RF} = 1$ ms at a power of -8 dBm; the solid line is a fit to a stretched exponential, see main text. (Bottom) Probability distribution for the diffusion rate $\mu$; the vertical dotted and dashed lines indicate the characteristic diffusion rate $1/\tau_d$ and distribution median $\bar{\mu}$. The shadowed half correspond to transport processes with rates faster than $\tau_d^{-1}$. (C) We model the observed response as a classical flow of magnetization through a chain of $m$ boxes, each containing $N_i$ spins with hyperfine resonance frequencies within box-selective-bandwidths $\Delta\nu_i$. The arrow indicates increasing hyperfine coupling $\|A\|$, and $\gamma_{i,i+1}$ denotes the polarization transfer rate between neighboring boxes. (D) Numerical simulations of the model in (C) for chains of length $m = 40$ and with uniform (but variable) spin transfer rate $\gamma$. We attain a sigmoidal response, whose inflection point at $\tau_d$ grows with the inverse of the spin diffusion rate $\gamma$. The magnetization contrast $\delta M_m$ reflects on the RF impact, here set to act on a fraction of the spins in the 20-th box of the chain. (E) Effective spin diffusion constant $D_{eff} = \langle r_C \rangle^2 \tau_d^{-1}$ at different RF frequencies $\nu_{RF}$ as determined from data plots similar to those in (C); $D_C$ is the spin diffusion for carbon in pure diamond. The broad green band reproduces the RF absorption from Fig. 2D and has been included as a reference. All experiments are carried out at a fixed magnetic field $B^{(+)} = 52.3$ mT.

Fig. 4D summarizes numerical results from a chain of $m = 40$ spin cells, each connected to its immediate neighbors via transfer rates $\gamma_{i,i+1}$. To establish a starting connection between $\tau_d$ and the underlying rates, we first investigate the case where $\gamma_{i,i+1}$ takes a constant value $\gamma$ throughout the chain (Fig. 4D). As expected, we find that $\tau_d$ grows inversely with $\gamma^{-1}$, though the dependence is not linear, a consequence of the finite duration of the RF pulse. To investigate the impact of transport heterogeneity, we also consider the case where $\gamma_{i,i+1}$ takes on different values depending on the position across the chain, peaking at the midpoint. Imposing greater transfer rates between cells effectively amounts to fusing neighboring groups of spins into a larger cell, hence amplifying the impact of individual RF pulses resonant with the set and thus altering $\tau_d$. Since the experimental response upon excitation at different frequencies does not substantially depart from our observations in Fig. 4B (see Supplementary Material, Section V), we tentatively conclude that the transfer rates across the hyperfine spectrum — or, more generally, the representative values from the transfer rate distributions connecting each cell in the chain with all others — are relatively uniform. On a related note, our numerical model exhibits only a moderate departure from a single exponential response ($\varepsilon \sim 0.9$ in Fig. 4D). The latter could well be a consequence of the first neighbor coupling structure assumed for the spin chain, likely oversimplifying the system complexity by limiting the number of channels available to the transport of nuclear polarization.

Capitalizing on the above considerations to interpret our observations, we find that the characteristic diffusion rate $\tau_d^{-1}(\nu_{RF})$ in the present spin system falls within the range 0.3-0.6 ms$^{-1}$. The agreement with the quantum model in Figs. 3E and 3F should be considered rather fortuitous as a numerical value of the diffusion time can only emerge from a suitable average over the set of possible spin configurations. On the other hand, given the mean inter-carbon distance in diamond $\langle r_C \rangle = 0.5$ nm, we conclude the effective diffusion constant observed herein can be as large as $D_{eff} \sim \langle r_C \rangle^2 \tau_d^{-1} \sim 1.5 \times 10^2$ nm$^2$ s$^{-1}$, about 100-fold greater than that derived from nuclear dipolar interactions alone[33,34] (Fig. 4C). This result reinforces the understanding of the cross-relaxing electron spin bath as a mediator to swiftly move around magnetization from otherwise many-body localized groups of nuclei. Such behavior could prove advantageous to expedite the transport of polarization across the diamond surface into arbitrary nuclear spin targets[16,35].

## DISCUSSION

While the effective Hamiltonian in Eq. (1) supports the notion of a coherent, electron-mediated nuclear spin transport, a question of interest is whether spin-lattice relaxation (electronic or nuclear) impacts the diffusion process itself (beyond imposing a limit on the polarization buildup). Supporting this notion, recent numerical studies suggest incoherent dynamics can help drive the spin system away from 'blockade' regimes, i.e., spin configurations that prevent the transport of spin polarization[35]. In the present framework, such processes could, e.g., flip P1 centers that have previously been polarized upon cross-relaxation with the NVs. Note that coherent channels remain the main transport driver and interacting paramagnetic defects are still central to the process, but here it is spin-lattice relaxation (not necessarily electron spin diffusion) that prepares the P1 for the next cycle of spin transport. This picture is consistent with the measured $\tau_d$, on average comparable to the NV or P1 spin lattice relaxation times (of order 1 ms in this diamond sample). Future experiments, for example, above and below room temperature or for samples with variable NV or P1 concentrations could help shed light on the role of incoherent processes.

Extensions of the ideas introduced herein can provide additional insights on the complex spin dynamics at play. For example, the use of chirped MW pulses to induce nuclear spin polarization[36,37] — away from the NV-P1 cross relaxation condition — can be exploited to separate the roles of NVs and P1s during the spin diffusion process. Along the same lines, microwave manipulation of the electron spin bath should give us the opportunity to controllably reintroduce localization in the nuclear spin system or to count the number of correlated carbons as the polarization spreads[38,39]. Particularly attractive is the combined use of super-resolution microscopy[40-42] and magnetic resonance techniques to monitor the spin dynamics of small ensembles of nuclear spins communicating via NV-P1 networks, which could be relevant to quantum information processing with many body disordered systems[43].

While our experiments centered on spins in diamond, we anticipate similar techniques can be adapted to investigate the dynamics of other material systems hosting spin active nuclear and electronic spins. These include organic systems exhibiting (non-optical) dynamic nuclear polarization, where simultaneous microwave and radio-frequency excitation could be exploited to gain information on nuclear spins proximal to radicals, normally invisible in standard DNP-enhanced NMR experiments.

## MATERIALS AND METHODS
### Experiment
Throughout our measurements we use a CVD-grown, type 1b diamond, which was previously electron-irradiated and annealed to create NV centers throughout the bulk crystal at an approximate concentration of 10 ppm[15]. Dynamic nuclear polarization is carried out via 1 W laser excitation at 532 nm. We rely on a pair of coils and the stray field of a 9 T NMR magnet to adjust the magnetic field during optical illumination in the vicinity of 51 mT, and use a home-made, compressed-air-driven device to shuttle the NMR probe between the polarization (51 mT) and detection fields (9 T). We orient the diamond crystal so that the external magnetic field nearly coincides with one of the NV axes, and use a single loop around the crystal as the source of RF excitation. All experiments are conducted at room temperature. Additional details, including a characterization of the $^{13}C$ spin response as a function of the RF power are presented in Section I of the Supplementary Material.

### Numerical simulations
We use a fourth order Trotter-Suzuki (TS) method[44] to numerically evaluate the time-dependence of large spin systems. The TS protocol avoids manipulating and diagonalizing the full Hamiltonian $H$, instead approximating the total evolution operator $U(t) = \exp\{-iHt\}$ by a suitable sequence of partial evolution operators $\widetilde{U}(\delta t) = \prod_k \exp\{-iH_k \delta t\}$. Here, $\{H_k\}$ corresponds to each of the single-spin and two-spin terms in the Hamiltonian $H$, proportional to either linear ($I_n^\alpha, \alpha = x, y, z$) or bilinear operators ($I_n^\alpha I_m^\alpha, \alpha = x, y, z$). The evaluation of the time-evolution for an arbitrary finite time $t$ requires the successive application of the steplike evolutions. Importantly, the approximated dynamics remains always unitary and the accuracy of the approximation relies on the TS time step $\delta t$ being sufficiently small as compared to the shortest local time scale of the original Hamiltonian. In our simulations, we have tuned the TS time step so that, for the system size considered (22 spins), relative errors bounds are estimated to be $10^{-4}$. Regarding the computational implementation, an exponential speedup of our simulations is achieved by means of a massive parallelization scheme via general-purpose graphical processing units[45] (GP-GPU).

**Acknowledgments**
**Funding.** D.P., J.H., and C.A.M. acknowledge support from the National Science Foundation through grants NSF-1903839 and NSF-1619896, and from Research Corporation for Science Advancement through a FRED Award; they also acknowledge access to the facilities and research infrastructure of the NSF CREST IDEALS, grant number NSF-HRD-1547830. J.H. acknowledges support from CREST-PRF NSF-HRD 1827037. All authors acknowledge the CUNY High Performance Computing Center (HPCC). The CUNY HPCC is operated by the College of Staten Island and funded, in part, by grants from the City of New York, State of New York, CUNY Research Foundation, and National Science Foundation Grants CNS-0958379, CNS-0855217 and ACI 1126113. **Author contributions:** C.A.M., D.P. and P.R.Z. conceived the experiment. D.P. and J.H. conducted the experiments. P.R.Z. developed a model and carried out theoretical calculations. A.A., R.A., J.A.R. and A.P. advised on the several aspects of theory and experiments. C.A.M. wrote the manuscript with input from all authors. R.A., J.A.R., A.P. and C.A.M. supervised the overall research effort. **Competing interests:** All authors declare that they have no competing interests. **Data and materials availability:** All data needed to evaluate the conclusions in the paper are present in the paper and/or the Supplementary Materials. Additional data related to this paper may be requested from the authors. All correspondence and request for materials should be addressed to C.A.M. (cmeriles@ccny.cuny.edu).

# Supporting Information

# *Optically pumped spin polarization as a probe of many-body thermalization*

Daniela Pagliero[1], Pablo R. Zangara[6,7], Jacob Henshaw[1], Ashok Ajoy[3], Rodolfo H. Acosta[6,7], Jeffrey A. Reimer[4,5], Alexander Pines[3,5], Carlos A. Meriles[1,2,*]

[1] Department. of Physics, CUNY-City College of New York, New York, NY 10031, USA. [2] CUNY-Graduate Center, New York, NY 10016, USA. [3] Department of Chemistry, University of California at Berkeley, Berkeley, California 94720, USA. [4] Department of Chemical and Biomolecular Engineering, University of California at Berkeley, Berkeley, California 94720, USA. [5] Materials Science Division Lawrence Berkeley National Laboratory, Berkeley, California 94720, USA. [6] Facultad de Matemática, Astronomía, Física y Computación, Universidad Nacional de Córdoba, Ciudad Universitaria, CP:X5000HUA Córdoba, Argentina. [7] Instituto de Física Enrique Gaviola (IFEG), CONICET, Medina Allende s/n, X5000HUA, Córdoba, Argentina.

## I.   Experimental

The experimental setup is a modified version of the system described in Ref. [15]. Briefly, it consists of a 400 MHz solid-state NMR magnet and spectrometer with a pneumatic shuttling device (Fig. S1A). During the hyperpolarization sequence, the sample is kept outside of the bore of the magnet, in the magnet's stray field, at about 52.3 mT. An electromagnet, with current provided by a programmable power supply (GW Instek PSM-6003) is used to fine tune the magnetic field to the hyperpolarization condition (Fig. S1B). The sample is optically pumped at low field with a 532 nm laser with ~ 700 mW at the sample. The laser is pulsed with an AOM (acousto-optic modulator, Isomet 1250C) for time-resolved measurements. The beam diameter is adjusted using a lens just before the sample. The pneumatic shuttling system sends the sample to the magnet's "sweet spot" in ~1s and a $^{13}$C FID is subsequently collected. The shuttling and spectrometer triggering are controlled with TTL pulses from a National Instruments DAQ card (PCIe 6321).

The NMR probe — which moves along with the sample — has been altered slightly from Ref. [15] to allow for manipulation of the $^{13}$C spins at low field in the hyperpolarization process. The RF is provided by an additional loop of wire near the sample. The loop terminates to either a 50-ohm resistor or shorts to ground to form a stub antenna depending on power needs. The RF signal is generated by a Rhode & Schwarz SMV 03 and amplified with a Minicircuits LZY-22+. Before amplification, the RF signal is gated by a switch (Minicircuits ZASWA-2-50DR+). Due to the bandwidth of the amplifier overlapping with the bandwidth of our spectrometer's receiver, the blanking control line of the amplifier is used to reduce the noise level in the detected signal. The RF amplifier blanking is controlled by the DAQ card. For experiments requiring precise time resolution, the AOM and the gate for the MW switch are controlled with pulses from a SpinCore Pulseblaster-300.

The RF power is calibrated by detecting $^{13}$C Rabi oscillations (Figs. S1C and S1D). To this end, a hyperpolarization step is performed; the laser pumps the diamond for 10 seconds with the magnetic field tuned close to 52.3 mT to where the hyperpolarization is maximum. At the end of the pumping, just before the shuttling, an RF pulse, resonant with the $^{13}$C Larmor frequency is applied. The sequence is repeated 4 times and averaged. This is done for a range of RF pulse durations allowing us to extract the Rabi frequency and hence, the $B_1$ magnetic field amplitude. An oscilloscope is used to monitor the peak-to-peak voltage. The output of the signal generator is adjusted so as to maintain the peak-to-peak voltage unchanged for all frequencies used.

Experiments to determine the impact of different hyperfine coupled $^{13}$Cs (Fig. 2 in the main text) were performed by tuning the magnetic field to the positive or negative hyperpolarization features associated with the P1's central Zeeman transition. For the present crystal orientation, this occurs slightly below or above 52.5 mT for the positive or negative polarizations (respectively $B^{(+)}$ and $B^{(-)}$, in Fig. 1D of the main text). The AOM and RF switch are triggered both at the same time for a variable duration, typically 5-10 s. This is repeated several times (normally 8) for each RF frequency. When the RF excites a hyperfine coupled carbon, this shorts the hyperpolarization diffusion process, lowering the hyperpolarization signal. To characterize nuclear spin diffusion (Fig. 4 in the main text), a pulse sequence consisting of 2 ms laser pulses separated by a dark time with variable duration RF pulses is looped until 2 seconds of laser time has been accumulated and averaged 12 times per RF pulse duration. A reference with the same dark time and no RF pulse is taken to compare the impact of the RF pulse.

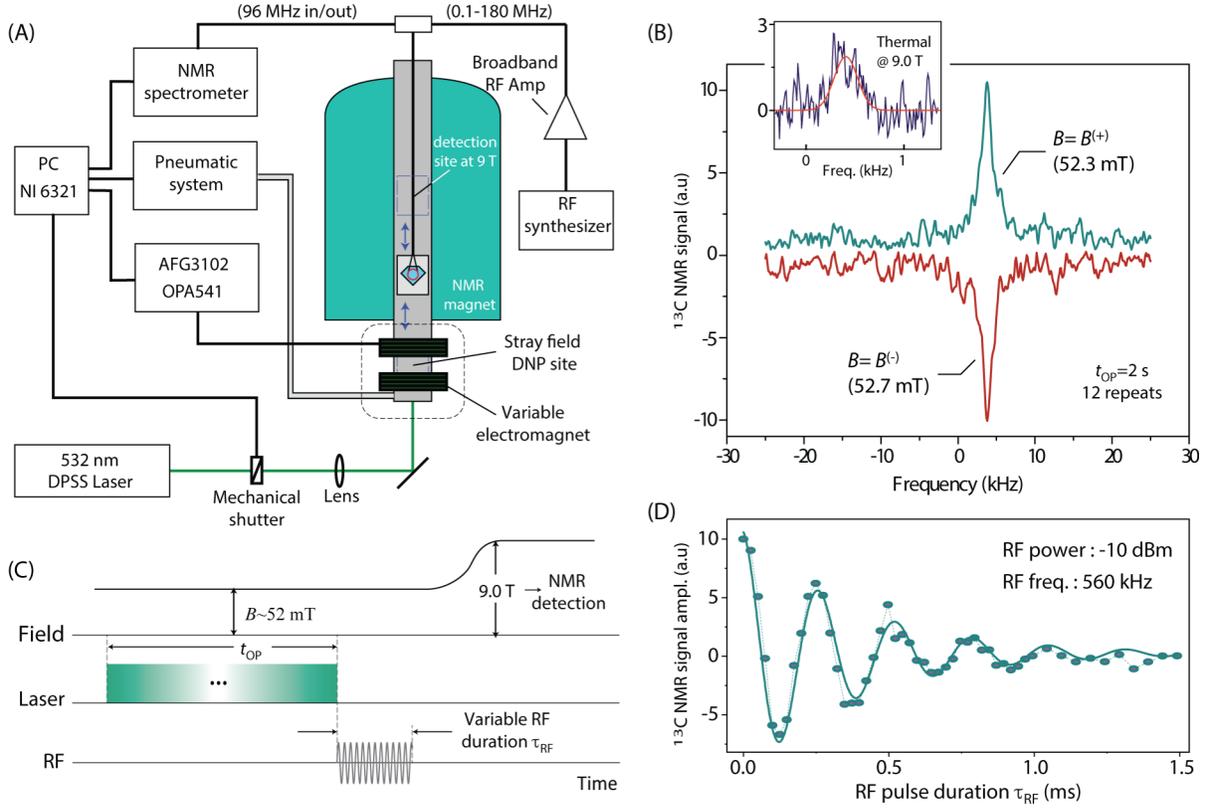

**Figure S1. Stray-field optical pumping of $^{13}$C polarization in diamond.** (A) Experimental setup. Dynamic polarization of $^{13}$C spins in diamond is carried out via optical excitation in the stray field of a superconducting NMR magnet (9.0 T). To detect the induced polarization, we shuttle the sample to the sweet spot of the magnet and acquire a free-induction-decay (FID) upon a resonant π/2-pulse at 96.87 MHz. (B) Example $^{13}$C NMR spectra upon 2-second optical excitation and a laser power of 0.7 W; the total number of repeats is 12 and spectra have been slightly displaced vertically for clarity. The signal is positive or negative depending on the exact value of the optical pumping field (see main text for details). From a comparison against the thermal signal amplitude at 9 T (24 repeats, upper inset) and the size of the illuminated spot (~100 μm), we conclude the dynamically pumped $^{13}$C polarization is of order 1-3%. (C) $^{13}$C spin manipulation at low field ($B^{(+)} = 52.3$ mT) is carried out via RF excitation at a variable frequency prior to sample shuttling. As an example, the cartoon shows the schematics of a Rabi protocol at low field. (D) $^{13}$C NMR amplitude as a function of the RF pulse duration as obtained from the protocol in (C). The polarization field is $B^{(+)} = 52.3$ mT and the RF frequency (560 MHz) is resonant with the bulk carbon Zeeman transition. The optical illumination time is 10 s and the number of repeats per point is 4.

## II. The spin Hamiltonian

The main goal in this and the following sections is to provide a quantum-mechanical model that qualitatively describes the polarization flow from strongly hyperfine-coupled $^{13}$Cs to bulk $^{13}$Cs. In order to simplify our formal description, we assume that the hyperfine-coupled carbons are initially polarized (by means of the NV-P1 energy matching mechanism discussed in Ref. [15]) and focus specifically on the spin-diffusion process. This means that we do not need to include the primary source of polarization, i.e. the NVs. The complete spin system therefore comprises $N_S$ electrons (P1 centers) and $N_I$ nuclear spins ($^{13}$C). The Hamiltonian describing this system is given by:

$$H_{\mathrm{T}} = \sum_{k,l}^{N_{\mathrm{I}}} H_{\mathrm{D}}(\boldsymbol{I}_k, \boldsymbol{I}_l) + \sum_{m,n}^{N_{\mathrm{S}}} H_{\mathrm{D}}(\boldsymbol{S}_m, \boldsymbol{S}_n) + \sum_{k}^{N_{\mathrm{I}}} \gamma_{\mathrm{I}} \boldsymbol{B} \cdot \boldsymbol{I}_k + \sum_{m}^{N_{\mathrm{S}}} \gamma_{\mathrm{S}} \boldsymbol{B} \cdot \boldsymbol{S}_m + \sum_{k}^{N_{\mathrm{I}}} \sum_{m}^{N_{\mathrm{S}}} V_{\mathrm{int}}(\boldsymbol{I}_k, \boldsymbol{S}_m) \qquad (\mathrm{A}.1)$$

Here, the first term corresponds to the dipolar interaction between $^{13}$C nuclear spins, the second is the dipolar interaction between P1 centers, the third and fourth are the corresponding Zeeman contributions, and the last term corresponds to the interaction between the two-spin species. Due to the typical $^{13}$C-$^{13}$C spatial separation in samples with natural $^{13}$C abundance, the first term corresponds to a very weak interaction, which we neglect.

The term involving $V_{\mathrm{int}}$ corresponds to the hyperfine couplings between electronic and nuclear spins. A direct flip-flop between a P1 spin and a $^{13}$C spin is not allowed due to the large energy mismatch $\gamma_{\mathrm{I}} B \ll |\gamma_{\mathrm{S}} B|$. Then, we are left with:

$$\sum_{k}^{N_{\mathrm{I}}} \sum_{m}^{N_{\mathrm{S}}} V_{\mathrm{int}}(\boldsymbol{I}_k, \boldsymbol{S}_m) \approx \sum_{k}^{N_{\mathrm{I}}} \sum_{m}^{N_{\mathrm{S}}} A_{zz}^{(m,k)} S_m^z I_k^z + A_{zx}^{(m,k)} S_m^z I_k^x \qquad (\mathrm{A}.2)$$

Notice here that the second term (known as pseudo-secular) cannot be truncated since, in the case of interest, the hyperfine energies exceed the nuclear Zeeman energy. For future reference, Eq. (A.1) can be easily extended to include an NV center provided the magnetic field is chosen so that the frequency of the $|0\rangle \leftrightarrow |-1\rangle$ NV transition matches the electron Larmor frequency, namely, when $|\gamma_{\mathrm{S}} B| \sim \Delta/2$, where $\Delta = 2.87$ GHz denotes the NV zero field splitting. This condition — met near 51 mT — immediately implies that the transfer of polarization from carbons coupled to an NV center is field dependent.

### III. The four-spin system and the effective $^{13}$C-$^{13}$C mechanisms

To analyze the dynamics induced by Eqns. (A.1) and (A.2), we start by considering a simple system with two $^{13}$Cs and two P1 centers, as shown in Fig. S2. Our objective is to derive an effective description of the dynamics of polarization within a spin system only composed by $^{13}$Cs. We start by writing down the Hamiltonian $H_{\mathrm{T}}$ in Eq. (A.1) for the model depicted in Fig. S2,

$$H_{\mathrm{T}} = -\omega_{\mathrm{I}} I_1^z - \omega_{\mathrm{I}} I_4^z + \omega_{\mathrm{S}} S_2^z + \omega_{\mathrm{S}} S_3^z + A_{zz}^{12} S_2^z I_1^z + A_{zx}^{12} S_2^z I_1^x + A_{zz}^{34} S_3^z I_4^z + A_{zx}^{34} S_3^z I_4^x + \mathcal{J}_{\mathrm{d}} \left( S_2^x S_3^x + S_2^y S_3^y \right) \qquad (\mathrm{A}.3).$$

Here, $\omega_{\mathrm{S}} = |\gamma_{\mathrm{S}} B|$, $\omega_{\mathrm{I}} = \gamma_{\mathrm{I}} B$ (note both frequencies are positive), and $\mathcal{J}_{\mathrm{d}}$ is the dipolar coupling between the two P1 centers (spins 2 and 3 in Fig. S2), and, as stated above, we assume an energy-matching external magnetic field $B = 51$ mT. Since $\omega_{\mathrm{S}}$ is the leading energy scale in $H_{\mathrm{T}}$, we can split it into three blocks given by the subspaces corresponding to P1-spin projection equal to 1, 0, -1. These blocks are not mixed by $H_{\mathrm{T}}$ since $[S_2^z + S_3^z, H_{\mathrm{T}}] = 0$. Furthermore, subspaces with spin projection 1 or -1 cannot yield an effective $^{13}$C-$^{13}$C interaction since the dynamics within these subspaces are equivalent to the evolution of the two $^{13}$Cs in the presence of an external static magnetic field. Thus, we restrict ourselves to the subspace of zero spin projection for the two P1 spins.

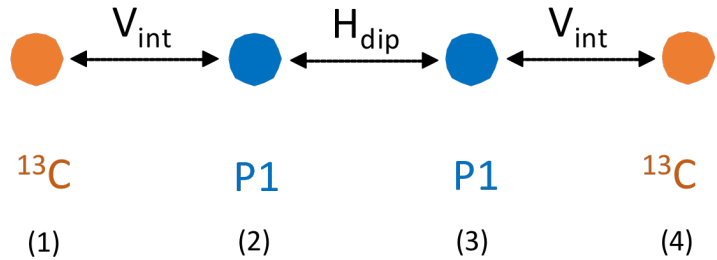

**Figure S2. The four-spin system.** Two $^{13}$Cs interact with two P1 centers. Each number labels the corresponding spin (see text).

By introducing the norm of the hyperfine interactions,

$$\Delta_{12} = \sqrt{(A_{zz}^{12})^2 + (A_{zx}^{12})^2} \tag{A.4}$$

$$\Delta_{34} = \sqrt{(A_{zz}^{34})^2 + (A_{zx}^{34})^2}. \tag{A.5}$$

we identify two different regimes defined by the hierarchy in the energy scales: Regime 1, where $\Delta_{12} \gtrsim \Delta_{34} > \mathcal{J}_d > \omega_I$, and Regime 2, where $\mathcal{J}_d > \Delta_{12} \sim \Delta_{34}, \omega_I$.

In what follows we analyze both regimes in detail.

*Regime 1. Hyperfine-dominated limit*

This case is characterized by $\Delta_{12} \gtrsim \Delta_{34} > \mathcal{J}_d > \omega_I$. Since the $^{13}$C quantization axis is essentially given by the hyperfine vector, it is natural to rewrite $H_T$ as:

$$H_T = -\omega_I I_1^z - \omega_I I_4^z + \omega_S S_2^z + \omega_S S_3^z + S_2^z(A_{zz}^{12}I_1^z + A_{zx}^{12}I_1^x) + S_3^z(A_{zz}^{34}I_4^z + A_{zx}^{34}I_4^x) + \mathcal{J}_d(S_2^x S_3^x + S_2^y S_3^y) \tag{A.6}$$

In order to diagonalize the hyperfine interaction, we rotate the local basis of each $^{13}$C spin to obtain

$$H_T = -\omega_z^{(1)} \tilde{I}_1^z + \omega_x^{(1)} \tilde{I}_1^x - \omega_z^{(4)} \tilde{I}_4^z + \omega_x^{(4)} \tilde{I}_4^x + \omega_S S_2^z + \omega_S S_3^z + \Delta_{12} S_2^z \tilde{I}_1^z + \Delta_{34} S_3^z \tilde{I}_4^z + \mathcal{J}_d(S_2^x S_3^x + S_2^y S_3^y) \tag{A.7}$$

where

$$\omega_z^{(1)} = \omega_I \frac{A_{zz}^{12}}{\Delta_{12}}$$

$$\omega_x^{(1)} = \omega_I \frac{A_{zx}^{12}}{\Delta_{12}}$$

$$\omega_z^{(4)} = \omega_I \frac{A_{zz}^{34}}{\Delta_{34}}$$

$$\omega_x^{(4)} = \omega_I \frac{A_{zx}^{34}}{\Delta_{34}}$$

Now we explicitly write down the Hamiltonian $H_T$ in the subspace of interest. Only for the purposes of simplifying the notation, we assume $\omega_z^{(1)} = \omega_z^{(4)} = \omega$ (in our simulations below, however, we lift this restriction and consider these values not necessarily equal).

|  | $\|\uparrow'\uparrow\downarrow\uparrow'\rangle$ | $\|\uparrow'\downarrow\uparrow\uparrow'\rangle$ | $\|\uparrow'\uparrow\downarrow\downarrow'\rangle$ | $\|\uparrow'\downarrow\uparrow\downarrow'\rangle$ | $\|\downarrow'\uparrow\downarrow\uparrow'\rangle$ | $\|\downarrow'\downarrow\uparrow\uparrow'\rangle$ | $\|\downarrow'\uparrow\downarrow\downarrow'\rangle$ | $\|\downarrow'\downarrow\uparrow\downarrow'\rangle$ |
|---|---|---|---|---|---|---|---|---|
| $\langle\uparrow'\uparrow\downarrow\uparrow'\|$ | $-\omega + \frac{\Delta_{12}-\Delta_{34}}{4}$ | $\mathcal{J}_d/2$ | $\omega_x^{(4)}/2$ | 0 | $\omega_x^{(1)}/2$ | 0 | 0 | 0 |
| $\langle\uparrow'\downarrow\uparrow\uparrow'\|$ | $\mathcal{J}_d/2$ | $-\omega + \frac{\Delta_{34}-\Delta_{12}}{4}$ | 0 | $\omega_x^{(4)}/2$ | 0 | $\omega_x^{(1)}/2$ | 0 | 0 |
| $\langle\uparrow'\uparrow\downarrow\downarrow'\|$ | $\omega_x^{(4)}/2$ | 0 | $\frac{\Delta_{12}+\Delta_{34}}{4}$ | $\mathcal{J}_d/2$ | 0 | 0 | $\omega_x^{(1)}/2$ | 0 |
| $\langle\uparrow'\downarrow\uparrow\downarrow'\|$ | 0 | $\omega_x^{(4)}/2$ | $\mathcal{J}_d/2$ | $\frac{-\Delta_{12}-\Delta_{34}}{4}$ | 0 | 0 | 0 | $\omega_x^{(1)}/2$ |
| $\langle\downarrow'\uparrow\downarrow\uparrow'\|$ | $\omega_x^{(1)}/2$ | 0 | 0 | 0 | $\frac{-\Delta_{12}-\Delta_{34}}{4}$ | $\mathcal{J}_d/2$ | $\omega_x^{(4)}/2$ | 0 |
| $\langle\downarrow'\downarrow\uparrow\uparrow'\|$ | 0 | $\omega_x^{(1)}/2$ | 0 | 0 | $\mathcal{J}_d/2$ | $\frac{\Delta_{12}+\Delta_{34}}{4}$ | 0 | $\omega_x^{(4)}/2$ |
| $\langle\downarrow'\uparrow\downarrow\downarrow'\|$ | 0 | 0 | $\omega_x^{(1)}/2$ | 0 | $\omega_x^{(4)}/2$ | 0 | $\omega + \frac{\Delta_{34}-\Delta_{12}}{4}$ | $\mathcal{J}_d/2$ |
| $\langle\downarrow'\downarrow\uparrow\downarrow'\|$ | 0 | 0 | 0 | $\omega_x^{(1)}/2$ | 0 | $\omega_x^{(4)}/2$ | $\mathcal{J}_d/2$ | $\omega + \frac{\Delta_{12}-\Delta_{34}}{4}$ |

(A.8)

Here, the prime in the $^{13}$C spin states indicates the quantization axis defined by the hyperfine vector. The two states highlighted in yellow and the two in green are nearly degenerate. If we focus on the green pair, i.e. states $|\uparrow'\uparrow\downarrow'\rangle$ and $|\downarrow'\downarrow\uparrow'\rangle$ (same argument valid for the pair $|\uparrow'\downarrow\uparrow\downarrow'\rangle$ and $|\downarrow'\uparrow\downarrow\uparrow'\rangle$)), second order perturbation theory yields a small energy difference $\delta$ that breaks the degeneracy,

$$\delta^{[1]} \approx 2\omega_x^{(1)}\omega_x^{(4)}\omega\frac{|\Delta_{12}^2 - \Delta_{34}^2|}{\Delta_{12}^2\Delta_{34}^2} \tag{A.9}$$

where the index 1 in square brackets stands for Regime 1. An effective description dealing only with $^{13}$C spins needs to incorporate a local field term accounting for this energy shift between the states $|\uparrow'\downarrow'\rangle$ and $|\downarrow'\uparrow'\rangle$.

Effective flip-flops can occur if we consider third-order processes,

$$|\uparrow'\downarrow\uparrow\downarrow'\rangle \to |\uparrow'\downarrow\uparrow\uparrow'\rangle \to |\uparrow'\uparrow\downarrow\uparrow'\rangle \to |\downarrow'\uparrow\downarrow\uparrow'\rangle$$

$$|\uparrow'\downarrow\uparrow\downarrow'\rangle \to |\downarrow'\downarrow\uparrow\downarrow'\rangle \to |\downarrow'\uparrow\downarrow\downarrow'\rangle \to |\downarrow'\uparrow\downarrow\uparrow'\rangle$$

and

$$|\uparrow'\uparrow\downarrow\downarrow'\rangle \to |\uparrow'\uparrow\downarrow\uparrow'\rangle \to |\uparrow'\downarrow\uparrow\uparrow'\rangle \to |\downarrow'\downarrow\uparrow\uparrow'\rangle$$

$$|\uparrow'\uparrow\downarrow\downarrow'\rangle \to |\downarrow'\uparrow\downarrow\downarrow'\rangle \to |\downarrow'\downarrow\uparrow\downarrow'\rangle \to |\downarrow'\downarrow\uparrow\uparrow'\rangle.$$

The sequences above include a single $^{13}$C spin-flip, a dipolar P1-P1 flip-flop, and finally a second single $^{13}$C spin flip. The entire process can be thought of as a virtual four-body interaction, already hinting at the effective mechanism of $^{13}$C-$^{13}$C flip-flop. More specifically, such a flip-flop occurs with a third-order coupling element

$$J_{\text{eff}}^{[1]} \approx \frac{4\omega_x^{(1)}\omega_x^{(4)}\mathcal{J}_d}{\Delta_{12}\Delta_{34}}. \tag{A.10}$$

Then, the proposed effective Hamiltonian describing the dynamics of the $^{13}$C pair in Regime 1 is:

$$H_{\text{eff}}^{[1]} = -\frac{\delta^{[1]}}{2}I_1^z + \frac{\delta^{[1]}}{2}I_4^z + J_{\text{eff}}^{[1]}(I_1^x I_4^x + I_1^y I_4^y) \tag{A.11}$$

We compare the dynamics induced by $H_T$ (Eq. (A.7)) and by $H_{\text{eff}}^{[1]}$ (Eq. (A.11)) in Fig. S3. In particular, we consider an initial state given by $|\uparrow'\downarrow\uparrow\downarrow'\rangle$ and monitor the time evolution of the polarization for both $^{13}$Cs. The comparison shows that the effective flip-flop mechanism can have a strength of up to a few kHz for strongly coupled P1 pairs. In fact, the flip-flop dynamics is dominant when $\delta^{[1]} < J_{\text{eff}}^{[1]} \propto \mathcal{J}_d$. Conversely, if the P1-P1 interaction is weak, then $\delta^{[1]} > J_{\text{eff}}^{[1]}$, and correspondingly the polarization remains *localized*. It is therefore natural to envision a direct generalization of $H_{\text{eff}}^{[1]}$ into the Anderson localization problem [9] for a large set of $N_I$ spins. In such case, the P1-P1 interaction controls the dynamical phase of the $^{13}$C system. In our simple two-spin case, a symbolic estimate for this localized-to-delocalized transition would occur at a critical interaction ($\delta^{[1]} \approx J_{\text{eff}}^{[1]}$)

$$\mathcal{J}_d^c \approx \omega\frac{|\Delta_{12}^2 - \Delta_{34}^2|}{2\Delta_{12}\Delta_{34}}. \tag{A.12}$$

We stress that an estimate of the mean P1-P1 interaction (or, equivalently, P1 concentration) needed to ensure spin diffusion within the $^{13}$C system requires a good knowledge of the statistical distribution of hyperfine couplings.

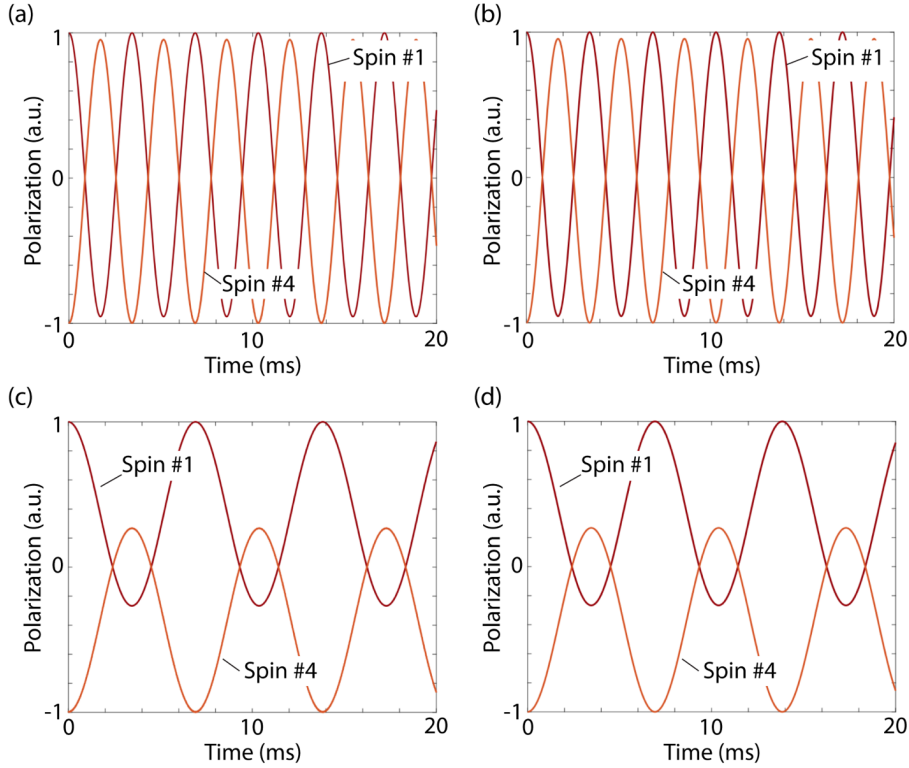

**Figure S3**. Comparison of the flip-flop dynamics (polarization transport) between the Hamiltonians $H_T$ (**a** and **c**) and $H_{\text{eff}}^{[1]}$ (**b** and **d**). In all the cases, $A_{zx}^{12} = A_{zz}^{12} = 40$ MHz, $A_{zx}^{34} = A_{zz}^{34} = 9$ MHz and the initial state is $|\uparrow'\uparrow\downarrow\downarrow'\rangle$. In (**a**) and (**b**), $J_d = 1$ MHz. In (**c**) and (**d**), $J_d = 5$ MHz.

We caution that the denominator in $J_{\text{eff}}^{[1]}$ (Eq. (A.10)) cannot be arbitrarily small. Further, the effective description also fails if there is a large mismatch between the hyperfine couplings (for example, when $\Delta_{12} \gg \Delta_{34} \sim \omega$). In such a case, the dynamics is essentially given by an uncorrelated single-spin flip at a frequency given by $\omega_x^{(4)}$,

$$|\uparrow'\uparrow\downarrow\downarrow'\rangle \leftrightarrow |\uparrow'\uparrow\downarrow\uparrow'\rangle$$

$$|\downarrow'\downarrow\uparrow\uparrow'\rangle \leftrightarrow |\downarrow'\downarrow\uparrow\downarrow'\rangle$$

$$|\uparrow'\downarrow\uparrow\downarrow'\rangle \leftrightarrow |\uparrow'\downarrow\uparrow\uparrow'\rangle$$

$$|\downarrow'\uparrow\downarrow\uparrow'\rangle \leftrightarrow |\downarrow'\uparrow\downarrow\downarrow'\rangle$$

and no polarization transfer happens (an equivalent example can be given for $\Delta_{34} \gg \Delta_{12} \sim \omega$). This scenario implies that strongly hyperfine-coupled $^{13}$C cannot transfer the polarization directly to bulk $^{13}$C. Nevertheless, strongly coupled $^{13}$C can effectively interact with 'moderately' hyperfine-coupled $^{13}$C, and these, in turn, interact with more weakly-hyperfine-coupled $^{13}$C, thus allowing the polarization to gradually cascade down to the bulk carbons.

Finally, by inspection of the matrix representation of $H_T$ in (A.8), it would be natural to expect terms of the form $I_1^z I_4^z$ in $H_{\text{eff}}^{[1]}$. These terms should account for the energy difference between the subspace spanned by $\{|\uparrow'\downarrow'\rangle, |\downarrow'\uparrow'\rangle\}$ and the subspace spanned by $\{|\uparrow'\uparrow'\rangle, |\downarrow'\downarrow'\rangle\}$. However, the simplified model employed herein is only useful to analyze the effective flip-flop mechanism and a more detailed analysis is required to derive the effective coupling element corresponding to an Ising term of the form $I_i^z I_j^z$. Such an approach, at the same time, would extend our previous discussion on (Anderson-) localization-delocalization into the many-body localization-delocalization problem.

*Regime 2. Dipolar-dominated limit*

This regime is characterized by $\mathcal{J}_d > \Delta_{12} \sim \Delta_{34}, \omega_I$, which means that the dipolar P1-P1 interaction defines the leading energy scale. We first consider the case $\mathcal{J}_d > \Delta_{12} \sim \Delta_{34} \gtrsim \omega_I$; notice, however, that since $\mathcal{J}_d$ does not largely exceed ~1 MHz for moderate P1 concentrations (50 ppm in the present case), both hyperfine interactions would have strengths comparable to $\omega_I$ (or, at least, not much higher than $\omega_I$), effectively limiting this regime to a narrow window. Then, we can choose here the $^{13}$C quantization axis to be given by the Zeeman interaction with the external magnetic field.

We further simplify this regime and assume, for now, $A^{12}_{zz} = A^{34}_{zz} = 0$. Then, the matrix representation of $H_T$ in Eq. (A.3) for the subspace of interest is

|  | $\|\uparrow\uparrow\downarrow\uparrow\rangle$ | $\|\uparrow\downarrow\uparrow\uparrow\rangle$ | $\|\uparrow\uparrow\downarrow\downarrow\rangle$ | $\|\uparrow\downarrow\uparrow\downarrow\rangle$ | $\|\downarrow\uparrow\downarrow\uparrow\rangle$ | $\|\downarrow\downarrow\uparrow\uparrow\rangle$ | $\|\downarrow\uparrow\downarrow\downarrow\rangle$ | $\|\downarrow\downarrow\uparrow\downarrow\rangle$ |
|---|---|---|---|---|---|---|---|---|
| $\langle\uparrow\uparrow\downarrow\uparrow\|$ | $-\omega_I$ | $\mathcal{J}_d/2$ | $-A^{34}_{zx}/4$ | 0 | $A^{12}_{zx}/4$ | 0 | 0 | 0 |
| $\langle\uparrow\downarrow\uparrow\uparrow\|$ | $\mathcal{J}_d/2$ | $-\omega_I$ | 0 | $A^{34}_{zx}/4$ | 0 | $-A^{12}_{zx}/4$ | 0 | 0 |
| $\langle\uparrow\uparrow\downarrow\downarrow\|$ | $-A^{34}_{zx}/4$ | 0 | 0 | $\mathcal{J}_d/2$ | 0 | 0 | $A^{12}_{zx}/4$ | 0 |
| $\langle\uparrow\downarrow\uparrow\downarrow\|$ | 0 | $A^{34}_{zx}/4$ | $\mathcal{J}_d/2$ | 0 | 0 | 0 | 0 | $-A^{12}_{zx}/4$ |
| $\langle\downarrow\uparrow\downarrow\uparrow\|$ | $A^{12}_{zx}/4$ | 0 | 0 | 0 | 0 | $\mathcal{J}_d/2$ | $-A^{34}_{zx}/4$ | 0 |
| $\langle\downarrow\downarrow\uparrow\uparrow\|$ | 0 | $-A^{12}_{zx}/4$ | 0 | 0 | $\mathcal{J}_d/2$ | 0 | 0 | $A^{34}_{zx}/4$ |
| $\langle\downarrow\uparrow\downarrow\downarrow\|$ | 0 | 0 | $A^{12}_{zx}/4$ | 0 | $-A^{34}_{zx}/4$ | 0 | $\omega_I$ | $\mathcal{J}_d/2$ |
| $\langle\downarrow\downarrow\uparrow\downarrow\|$ | 0 | 0 | 0 | $-A^{12}_{zx}/4$ | 0 | $A^{34}_{zx}/4$ | $\mathcal{J}_d/2$ | $\omega_I$ |

(A.13)

Now we transform the Hamiltonian into a basis that diagonalizes the dipolar P1-P1 interaction $\mathcal{J}_d(S^x_2 S^x_3 + S^y_2 S^y_3)$. Here, the eigenstates are

$$|+\rangle = \frac{1}{\sqrt{2}}(|\uparrow\downarrow\rangle + |\downarrow\uparrow\rangle) \tag{A.14}$$

$$|-\rangle = \frac{1}{\sqrt{2}}(|\uparrow\downarrow\rangle - |\downarrow\uparrow\rangle) \tag{A.15}$$

Then, the Hamiltonian matrix is given by

|  | $\|\uparrow -\uparrow\rangle$ | $\|\uparrow +\uparrow\rangle$ | $\|\uparrow -\downarrow\rangle$ | $\|\uparrow +\downarrow\rangle$ | $\|\downarrow -\uparrow\rangle$ | $\|\downarrow +\uparrow\rangle$ | $\|\downarrow -\downarrow\rangle$ | $\|\downarrow +\downarrow\rangle$ |
|---|---|---|---|---|---|---|---|---|
| $\langle\uparrow -\uparrow\|$ | $-\omega_I - \mathcal{J}_d/2$ | 0 | 0 | $-A^{34}_{zx}/4$ | 0 | $A^{12}_{zx}/4$ | 0 | 0 |
| $\langle\uparrow +\uparrow\|$ | 0 | $-\omega_I + \mathcal{J}_d/2$ | $-A^{34}_{zx}/4$ | 0 | $A^{12}_{zx}/4$ | 0 | 0 | 0 |
| $\langle\uparrow -\downarrow\|$ | 0 | $-A^{34}_{zx}/4$ | $-\mathcal{J}_d/2$ | 0 | 0 | 0 | 0 | $A^{12}_{zx}/4$ |
| $\langle\uparrow +\downarrow\|$ | $-A^{34}_{zx}/4$ | 0 | 0 | $\mathcal{J}_d/2$ | 0 | 0 | $A^{12}_{zx}/4$ | 0 |
| $\langle\downarrow -\uparrow\|$ | 0 | $A^{12}_{zx}/4$ | 0 | 0 | $-\mathcal{J}_d/2$ | 0 | 0 | $-A^{34}_{zx}/4$ |
| $\langle\downarrow +\uparrow\|$ | $A^{12}_{zx}/4$ | 0 | 0 | 0 | 0 | $\mathcal{J}_d/2$ | $-A^{34}_{zx}/4$ | 0 |
| $\langle\downarrow -\downarrow\|$ | 0 | 0 | 0 | $A^{12}_{zx}/4$ | 0 | $-A^{34}_{zx}/4$ | $\omega_I - \mathcal{J}_d/2$ | 0 |
| $\langle\downarrow +\downarrow\|$ | 0 | 0 | $A^{12}_{zx}/4$ | 0 | $-A^{34}_{zx}/4$ | 0 | 0 | $\omega_I + \mathcal{J}_d/2$ |

(A.16)

The subspace highlighted in green contains two pairs of quasi-degenerate states: $|\uparrow -\downarrow\rangle$ and $|\downarrow -\uparrow\rangle$ with energy $-\mathcal{J}_d/2$, and $|\uparrow +\downarrow\rangle$ and $|\downarrow +\uparrow\rangle$ with energy $+\mathcal{J}_d/2$. As before, second order perturbation theory provides an estimate for the energy shift that breaks degeneracy,

$$\delta^{[2]} \approx \omega_I \frac{|(A_{zx}^{34})^2 - (A_{zx}^{12})^2|}{8\mathcal{J}_d^2} \tag{A.17}$$

where the index 2 in square brackets refers to Regime 2. Again, an effective description only dealing with $^{13}$C spins needs to incorporate a local field term accounting for $\delta^{[2]}$.

An effective flip-flop mechanism can be derived also from second order processes,

$$|\uparrow +\downarrow\rangle \rightarrow |\uparrow -\uparrow\rangle \rightarrow |\downarrow +\uparrow\rangle$$

$$|\uparrow +\downarrow\rangle \rightarrow |\downarrow -\downarrow\rangle \rightarrow |\downarrow +\uparrow\rangle$$

and

$$|\uparrow -\downarrow\rangle \rightarrow |\uparrow +\uparrow\rangle \rightarrow |\downarrow -\uparrow\rangle$$

$$|\uparrow -\downarrow\rangle \rightarrow |\downarrow +\downarrow\rangle \rightarrow |\downarrow -\uparrow\rangle.$$

This type of transition involves a sequence of two $^{13}$C spin flips mediated by a virtual change in the P1 interaction energy. The obtained effective flip-flop mechanism has a coupling element

$$J_{eff}^{[2]} \approx \frac{A_{zx}^{34} A_{zx}^{12}}{4\mathcal{J}_d}. \tag{A.18}$$

where we use $\mathcal{J}_d > \omega_I$ to drop the dependence of $\omega_I$ in the denominator. This leads us to propose the following effective Hamiltonian,

$$H_{eff}^{[2]} = -\frac{\delta^{[2]}}{2} I_1^z + \frac{\delta^{[2]}}{2} I_4^z + J_{eff}^{[2]}(I_1^x I_4^x + I_1^y I_4^y), \tag{A.19}$$

From Eqns. (A.17) and (A.18), it is straightforward to verify that

$$\delta^{[2]} \lesssim \frac{\omega_I}{\mathcal{J}_d} J_{eff}^{[2]} < J_{eff}^{[2]}, \tag{A.20}$$

thus the polarization dynamics in Regime 2 is always delocalized.

We illustrate the accuracy of $H_{eff}^{[2]}$ by comparing the polarization dynamics induced by Eqns. (A.3) and (A.19). In Fig. S4 we consider an initial state given by $|\uparrow\downarrow\uparrow\downarrow\rangle$ and monitor the time evolution of the polarization for both $^{13}$Cs using the complete Hamiltonian $H_T$ and the effective $H_{eff}^{[2]}$. The comparison shows that the effective flip-flop mechanism can have a strength of hundreds of kHz, though only within the narrow window where $\mathcal{J}_d \sim A_{zx}^{12} \sim A_{zx}^{34} \gtrsim \omega_I$. Beyond this condition (i.e., when $\mathcal{J}_d \sim \omega_I$) we expect the effective Hamiltonian to gradually deviate from the exact $H_T$, since in Eq. (A.18) we disregard the effect of $\omega_I$ in the denominator of the coupling parameter. This situation is seen in Fig. S4(**a**) (where $\mathcal{J}_d = A_{zx}^{12} = 1$ MHz, $A_{zx}^{34} = 750$ kHz, and $\omega_I = \gamma_I B \approx 500$ kHz). The modulations present in the dynamics of the effective flip-flop Hamiltonian $H_{eff}$ can be more important if $A_{zz}^{12}, A_{zz}^{34} \neq 0$ because these terms contribute to the diagonal Hamiltonian matrix elements. Note that as long as $\mathcal{J}_d$ remains the leading energy scale, Eq. (A.18) applies beyond the condition $\mathcal{J}_d > \Delta_{12} \sim \Delta_{34} \gtrsim \omega_I$ to include the limit where the hyperfine shifts go to zero, i.e. $\mathcal{J}_d > \omega_I > \Delta_{12} \sim \Delta_{34}$.

A complete hierarchical picture of the $^{13}$C-$^{13}$C interactions can thus be drawn: Medium strength (few kHz) effective interactions develop inside the classical "diffusion barrier", provided the frequency mismatch between hyperfine couplings is sufficiently small (Regime 1); stronger effective interactions (reaching up to 100 kHz) become possible as carbons occupy positions farther removed from the electron spins, to subsequently decay as the hyperfine couplings gradually vanish (Regime 2). In this latter limit, carbon couplings take the value corresponding to that defined by the dipolar spin coupling between bulk nuclei.

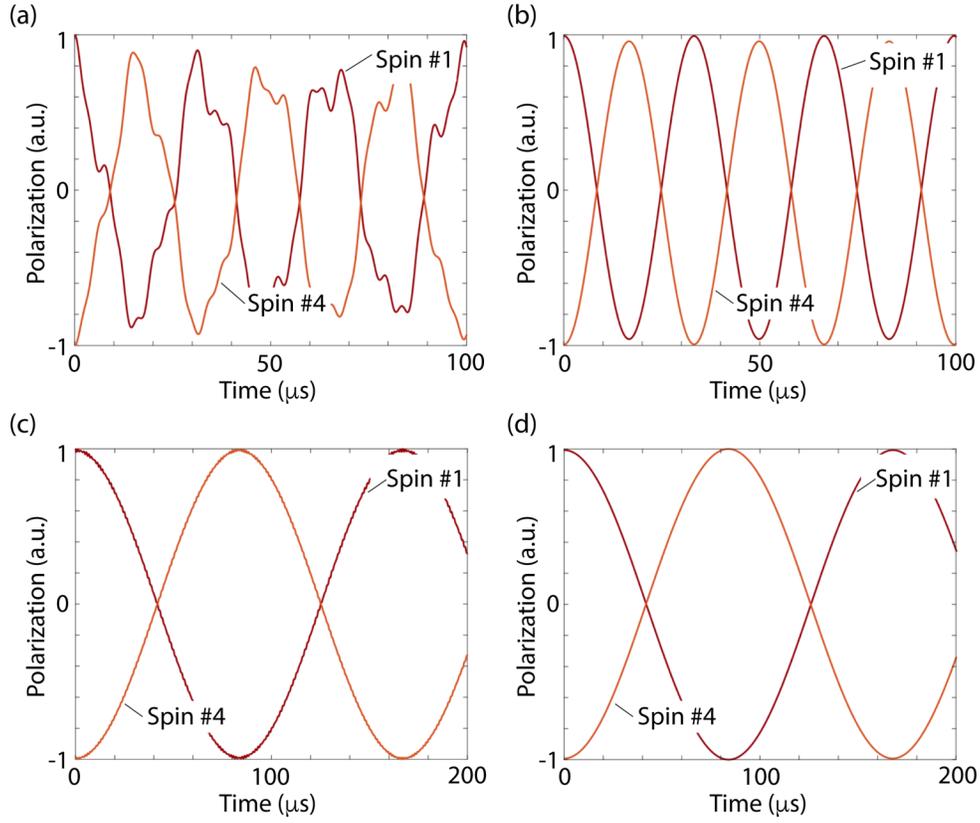

**Figure S4**. Comparison of the flip-flop dynamics (polarization) between the Hamiltonians $H_T$ (**a** and **c**) and $H_{eff}$ (**b** and **d**). In all cases, $A_{zx}^{12} = 1$ MHz, $A_{zx}^{34} = 750$ kHz, and the initial state is $|\uparrow\downarrow\uparrow\downarrow\rangle$. In (**a**) and (**b**), $\mathcal{J}_d = 1$ MHz. In (**c**) and (**d**), $\mathcal{J}_d = 5$ MHz.

## IV. The effect of RF excitation

To study the impact of RF on the system dynamics, we go back to the four-spin model and rewrite the Hamiltonian in Eq. (A.6) as,

$$H_T = -\omega_I I_1^z - \omega_I I_4^z + \omega_S S_2^z + \omega_S S_3^z + S_2^z(A_{zz}^{12} I_1^z + A_{zx}^{12} I_1^x) + S_3^z(A_{zz}^{34} I_4^z + A_{zx}^{34} I_4^x) + \mathcal{J}_d(S_2^x S_3^x + S_2^y S_3^y)$$
$$+ (I_1^x + I_4^x)\Omega\cos(\omega_{rf} t) \quad (A.21)$$

Now, assuming for concreteness Regime 1 and transforming into the hyperfine basis, we obtain

$$H_T = -\omega_z^{(1)} \tilde{I}_1^z + \omega_x^{(1)} \tilde{I}_1^x - \omega_z^{(4)} \tilde{I}_4^z + \omega_x^{(4)} \tilde{I}_4^x + \omega_S S_2^z + \omega_S S_3^z + \Delta_{12} S_2^z \tilde{I}_1^z + \Delta_{34} S_3^z \tilde{I}_4^z + \mathcal{J}_d(S_2^x S_3^x + S_2^y S_3^y)$$
$$+ \left(\Omega_z^{(1)} \tilde{I}_1^z + \Omega_x^{(1)} \tilde{I}_1^x\right)\cos(\omega_{rf} t) + \left(\Omega_z^{(4)} \tilde{I}_4^z + \Omega_x^{(4)} \tilde{I}_4^x\right)\cos(\omega_{rf} t) \quad (A.22)$$

where

$$\Omega_z^{(1)} = \Omega \frac{A_{zz}^{12}}{\Delta_{12}}$$

$$\Omega_x^{(1)} = \Omega \frac{A_{zx}^{12}}{\Delta_{12}}$$

$$\Omega_z^{(4)} = \Omega \frac{A_{zz}^{34}}{\Delta_{34}}$$

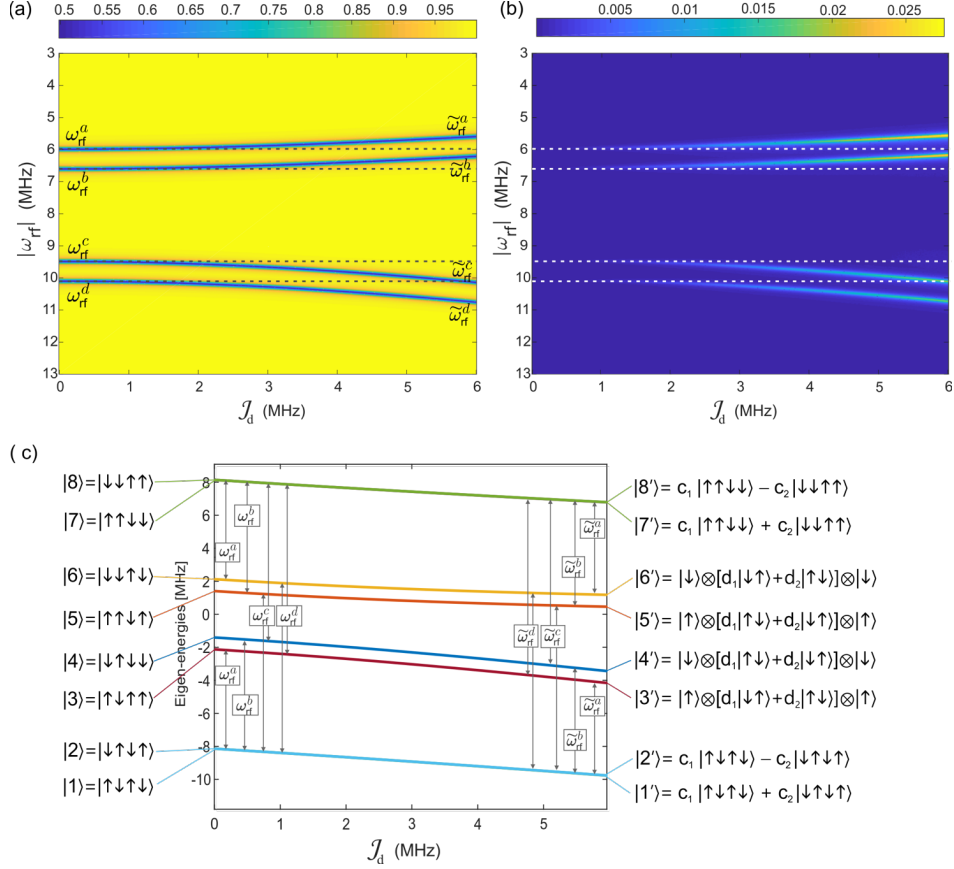

**Figure S5. Dynamical response in the presence of RF excitation.** (a) Polarization of $^{13}$C spins as a function of the P1-P1 dipolar interaction $\mathcal{J}_d$ and RF excitation $\omega_{rf}$. Both carbon spins are initially polarized. (b) P1 polarization (initially assumed to be zero) as a function of $\mathcal{J}_d$ and $\omega_{rf}$. (c) Eigen-energies as a function of $\mathcal{J}_d$ in the subspace of zero spin projection for the P1 spins. The same resonance frequencies shown in (a,b) are shown here with arrows. For the eigenstates in the regime of large P1-P1 interaction (e.g., $\mathcal{J}_d \sim 5$ MHz), the coefficients verify $c_1 \approx c_2$ and $d_1 \gg d_2$. In all cases we assume $A_{zz}^{12} = A_{zx}^{12} = 14$ MHz, $A_{zz}^{34} = A_{zx}^{34} = 9$ MHz, $B = 51$ mT. For (a) and (b), $\Omega = 75$ kHz.

$$\Omega_x^{(4)} = \Omega \frac{A_{zx}^{34}}{\Delta_{34}}$$

In the rotating frame, after performing standard time averaging, we finally write

$$H_T = \left(-\omega_z^{(1)} + \Omega_z^{(1)} - \omega_{rf}\right)\tilde{I}_1^z + \Delta_{12} S_2^z \tilde{I}_1^z + \left(-\omega_z^{(4)} + \Omega_z^{(4)} - \omega_{rf}\right)\tilde{I}_4^z + \Delta_{34} S_3^z \tilde{I}_4^z$$
$$+ \Omega_x^{(1)} \tilde{I}_1^x + \Omega_x^{(4)} \tilde{I}_4^x + \omega_S S_2^z + \omega_S S_3^z + \mathcal{J}_d\left(S_2^x S_3^x + S_2^y S_3^y\right) \quad \text{(A.23)}$$

To highlight the 'hybrid' electron/nuclear-spin nature of the transitions, here we assume both $^{13}$Cs are polarized and the P1 pair is in the subspace of zero spin projection (note that this is in contrast with the case in the main text where we assume both P1s are unpolarized). Fig. S5(**a-b**) shows both the nuclear and electronic polarization as a function of the excitation frequency $\omega_{rf}$ and the P1-P1 coupling parameter $\mathcal{J}_d$. To help understand these results, Figure S5(**c**) shows the energy spectrum as a function of $\mathcal{J}_d$ with an identification of the eigenstates in the two extreme cases $\mathcal{J}_d = 0$ (left) and $\mathcal{J}_d \sim 5$ MHz (right). In the limit of $\mathcal{J}_d = 0$, four possible transitions can be identified, which correspond to each $^{13}$C flipping independently. As $\mathcal{J}_d$ increases, the resonance frequencies are shifted and the states involved in each transition change accordingly. As the eigenstates feature contributions from different electron and nuclear spin projections, all transitions in this regime involve simultaneous nuclear and electronic spin-flips.

## V. Master equation approach: Spectral chain

This section of the Supplementary Material expands on the results of Fig. 4B in the main text, namely the response of the $^{13}$C NMR signal upon application of a train of RF pulse of variable separation simultaneous with optical illumination. Experimental results at various excitation frequencies along their inverse Laplace transforms are presented in Fig. S6. While a full quantum mechanical model is impractical, we can employ a classical master equation approach to analyze the magnetization flow from strongly hyperfine-coupled $^{13}$Cs to bulk $^{13}$Cs. The physical picture is based on a one-dimensional chain, where each link can be viewed as a spin set with a specific spectral location (hyperfine shift), as shown in Fig. 4C of the main text. More precisely, the magnetization charge $\{q_i\}_{i=1}^{m}$ of each of these boxes is described by

$$\frac{d}{dt}q_1 = -\gamma_{12}q_1 + \gamma_{21}q_2 - \beta_1 q_1 \tag{A.24}$$

$$\frac{d}{dt}q_2 = -\gamma_{21}q_2 - \gamma_{23}q_2 + \gamma_{12}q_1 + \gamma_{32}q_3 - \beta_2 q_2 \tag{A.25}$$

$$\frac{d}{dt}q_3 = -\gamma_{32}q_3 - \gamma_{34}q_3 + \gamma_{23}q_2 + \gamma_{43}q_4 - \beta_3 q_3 \tag{A.26}$$

/.../

$$\frac{d}{dt}q_k = -\gamma_{k,k-1}q_k - \gamma_{k,k+1}q_k + \gamma_{k-1,k}q_{k-1} + \gamma_{k+1,k}q_{k+1} - \beta_k q_k - a_{\text{RF}}g(t)\, q_k \tag{A.27}$$

/.../

$$\frac{d}{dt}q_m = \gamma_{m-1,m}q_{m-1} - \beta_m q_m \tag{A.28}$$

Here, $\gamma_{ij}$ stands for the transfer rate from box $i$ to box $j$, and $\beta_i$ represents the loss of magnetization due to nuclear spin-lattice relaxation. The RF excitation is resonant with box $k$, $g(t)$ stands for the shape of the train of RF pulses, and $a_{\text{RF}}$ is the amplitude of each pulse (here seen to act as a polarization sink). Consistent with the relative spectral proximity required for electron-spin-mediated transport (see Eqs. (A.10) and (A.12)), we only consider interactions between immediate spectral neighbors (i.e., $k-1$ and $k+1$) though additional contributions from farther removed boxes can be easily incorporated. Further, we neglect any backflow from the last box to the rest of the chain, and ignore non-linear (i.e., 'blockade') effects arising from saturation of the magnetization in a given box; this latter regime can always be attained when the illumination power is sufficiently low. The set of equations can then be written in the standard matrix form

$$\frac{d}{dt}\boldsymbol{Q} = \boldsymbol{A}\boldsymbol{Q}, \tag{A.29}$$

with

$$\boldsymbol{A} = \begin{pmatrix} -\gamma_{12} - \beta_1 & \gamma_{21} & 0 & 0 & \cdots & \cdots & 0 \\ \gamma_{12} & -\gamma_{21} - \gamma_{23} - \beta_2 & \gamma_{32} & 0 & 0 & 0 & 0 \\ 0 & \gamma_{23} & \cdots & 0 & 0 & 0 & 0 \\ 0 & 0 & \cdots & \gamma_{k,k-1} & 0 & 0 & 0 \\ \cdots & \cdots & \gamma_{k-1,k} & -\gamma_{k,k-1} - \gamma_{k,k+1} - \beta_k - a_{\text{RF}}g(t) & \gamma_{k+1,k} & 0 & 0 \\ 0 & 0 & 0 & \gamma_{k,k+1} & \cdots & \cdots & \cdots \\ 0 & 0 & 0 & 0 & \cdots & \cdots & \cdots \\ 0 & 0 & 0 & 0 & \cdots & \cdots & 0 \\ 0 & 0 & 0 & \cdots & 0 & \gamma_{m-1,m} & -\beta_m \end{pmatrix}.$$

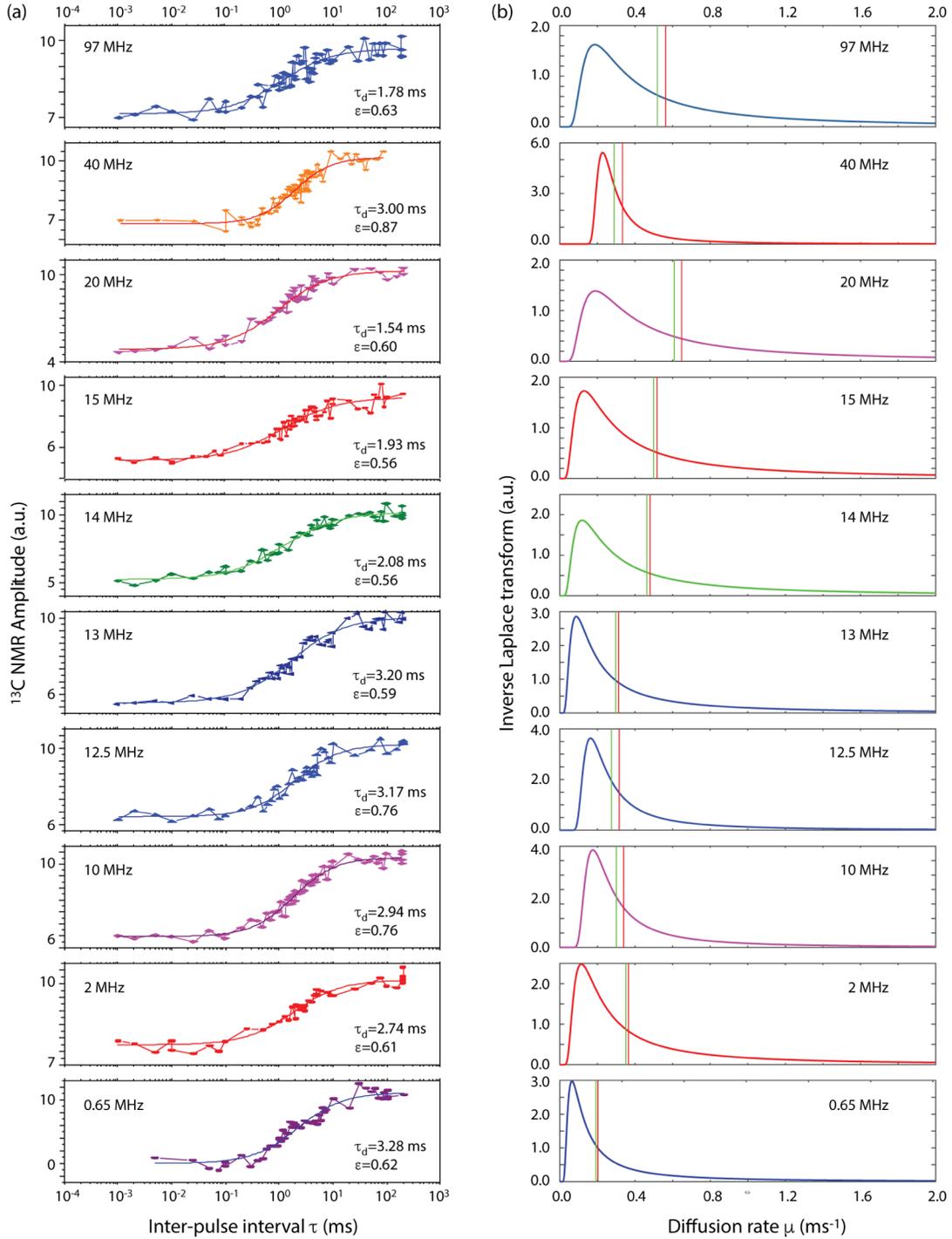

**Figure S6. Probing the time scale of $^{13}$C spin diffusion.** (**a**) We use the protocol in Fig. 4a of the main text to identify the effective nuclear spin diffusion time $\tau_d$ upon pulsed excitation at various frequencies (upper left corner in each plot). Solid lines represent fits to the stretched exponential function $S = S_0 - S_1 \exp(-(\tau/\tau_d)^\varepsilon)$, where $\tau_d$ is the characteristic nuclear spin diffusion time and $\varepsilon$, $S_0$, and $S_1$ are additional fitting parameters. (**b**) Laplace transforms of the stretched exponentials on the left. In each case, the vertical dashed and dotted lines indicate the distribution median and fitted value of $\tau_d$.

It is natural to split the evolution into intervals with and without RF excitation, since these correspond to $A_1 \equiv A(g=1)$ and $A_0 \equiv A(g=0)$, respectively. Using $\tau$ to denote the inter-pulse delay and $\tau_{RF}$ to indicate the RF pulse duration (here fixed to 1ms), the evolution of the magnetizations in each composite interval is given by

$$Q(\tau + \tau_{RF}) = \exp(A_0\tau) \exp(A_1\tau_{RF}) Q_0. \qquad (A.30)$$

Given a total evolution time $T$, the number of composite intervals is given by $n_p = T/(\tau + \tau_{RF})$. Then, the final magnetization is given by

$$Q(T) = [\exp(A_0\tau) \exp(A_1\tau_{RF})]^{n_p} Q_0. \qquad (A.31)$$

In our simulations, we consider an initial condition given by $q_1 = 1$ and $q_i = 0$ $\forall i > 1$. This is a crude approximation since we do not include the continuous effect of the optical pumping. Additionally, we also assume for simplicity $\beta_i = 0$ $\forall i$. The intensity of the RF irradiation is the leading scale of the problem, here assumed to be $a_{RF} = 1$ MHz. The total time considered is always $T = 1$ s.

The first case we study corresponds to a uniform set of coupling constants, $\gamma_{ij} = \gamma_{ji} = \gamma_0$. Figure S7 shows the normalized magnetization charge in the last box after the full evolution $q_m(T = 1 \text{ s})$ as a function of the inter-pulse delay time $\tau$, for different $\gamma_0$. The system has $m = 40$ boxes and the RF-irradiated box is always $k = 20$.

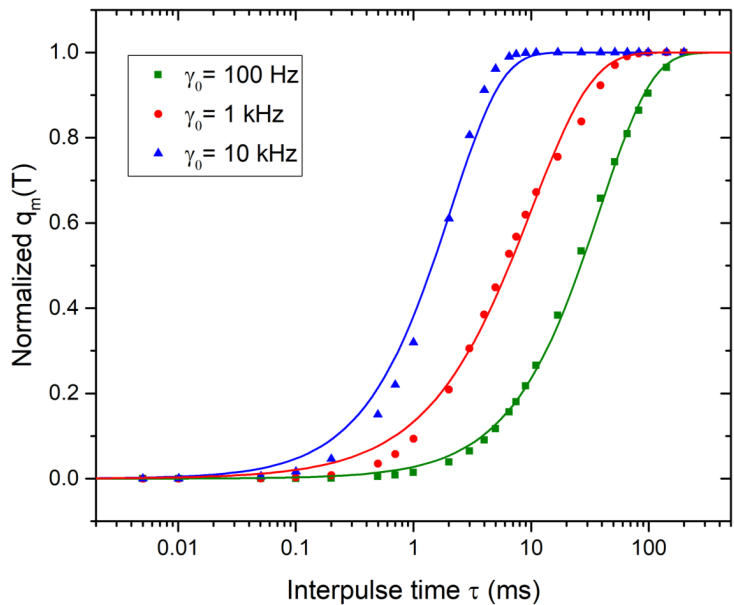

**Figure S7.** Normalized magnetization in the end box ($m$ − th) in a system of 40 boxes upon RF irradiation in box $k = 20$. The couplings constants are uniformly distributed, $\gamma_{ij} = \gamma_{ji} = \gamma_0$, with $\gamma_0$ given in the inset. The solid lines are given by stretched-exponential fittings, with $\varepsilon = 1$ for the blue and green cases, and $\varepsilon = 0.8$ for the red case.

In Fig. S8 we investigate the dependence of $q_m(T = 1 \text{ s})$ on the point of RF excitation across the chain. In the case of uniform couplings (Fig. S8a), we verify that the observed time-scale does not depend on the location of the saturated box. In Fig. S8b we consider a small, localized fraction of the chain has much stronger couplings than the rest. In particular, we assume the coupling set given by

$$\gamma_{i,i+1} = \gamma_{i+1,i} = \gamma_0 + 100\gamma_0 \exp\left\{-\left(\frac{k_0 - i}{K_0}\right)^2\right\}, \qquad (A.32)$$

where we choose $k_0 = 15$ and $K_0 = 2$. In this case, we observe a stronger attenuation of $q_m(T = 1 \text{ s})$ as we irradiate the boxes close to the box $k_0$. This means that saturating strongly connected nodes produces a stronger degradation in the magnetization reaching the end of the chain.

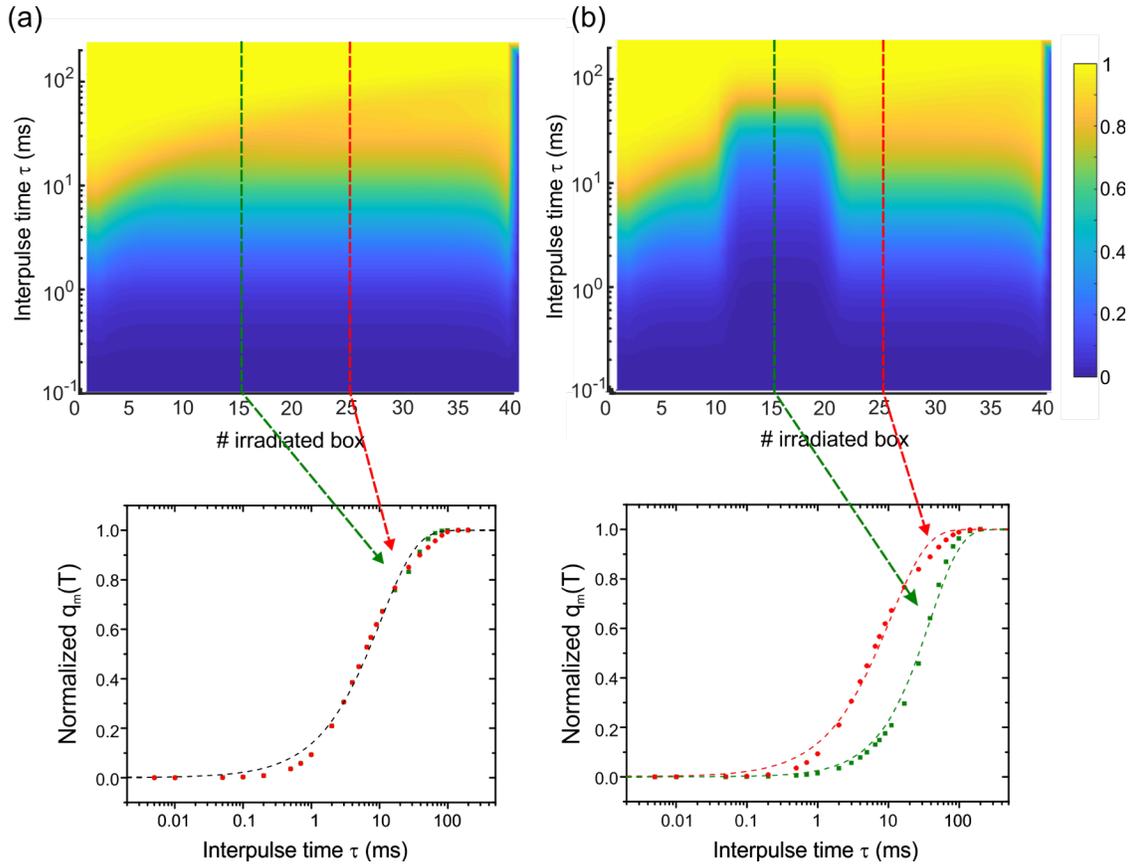

**Figure S8.** Normalized magnetization in the end box of a chain of 40 upon RF irradiation. Lower panels explicitly show when boxes 15 and 25 are being (independently) irradiated (dashed lines are stretched exponential fittings, with $\varepsilon = 0.8$. In case (**a**), the couplings constants are uniformly distributed, $\gamma_{ij} = \gamma_{ji} = \gamma_0$, with $\gamma_0 = 1$ kHz. In (**b**) we consider a uniform distribution perturbed in a small region around $k_0 = 15$ where the couplings $\gamma_{i,i+1}$ can be up to 100 times $\gamma_0$.